\documentclass[10pt,preprint]{aastex}

\usepackage{rotating}

\slugcomment{accepted by ApJ}

\shorttitle{Chemical Enrichment of the Antennae hot ISM II}
\shortauthors{Baldi et al.}

\begin{document}

\title{Chemical enrichment of the complex hot ISM of the Antennae Galaxies: II. Physical
properties of the hot gas and supernova feedback}


\author{A. Baldi, J.C. Raymond, G. Fabbiano, A. Zezas, A.H. Rots} 
\affil{Harvard-Smithsonian Center for Astrophysics, 60 Garden St, Cambridge, MA 
02138}
\email{abaldi@cfa.harvard.edu; jraymond@cfa.harvard.edu; pepi@cfa.harvard.edu; 
azezas@cfa.harvard.edu; arots@cfa.harvard.edu}

\author{F. Schweizer}
\affil{Carnegie Observatories, 813 Santa Barbara St, Pasadena, CA 91101}
\email{schweizer@ociw.edu}

\author{A.R. King}
\affil{Theoretical Astrophysics Group, University of Leicester, Leicester LE1
7RH, UK}
\email{ark@astro.le.ac.uk}

\and

\author{T.J. Ponman}
\affil{School of Physics \& Astronomy, University of Birmingham, Birmingham 
B15 2TT, UK}
\email{tjp@star.sr.bham.ac.uk}
%
%

\begin{abstract}
We investigate the physical properties of the interstellar medium (ISM) in
the merging pair of galaxies known as The Antennae (NGC 4038/39), using the deep
coadded $\sim$411 ks {\it Chandra} ACIS-S data set. The method of analysis and
some of the main results from the spectral analysis, such as metal abundances
and their variations from $\sim$0.2 to $\sim$20--30 times solar, are described in
Paper I (Baldi et al. submitted).
In the present paper we investigate in detail the physics of the hot emitting gas,
deriving measures for the hot-gas mass ($\sim$10$^7 M_\sun$), cooling times 
($10^7$--$10^8$ yr), and pressure ($3.5\times10^{-11}$--$2.8\times10^{-10}$
dyne cm$^{-2}$). At least in one of the two nuclei (NGC 4038) the hot-gas 
pressure is significantly higher than the CO pressure, implying that shock waves 
may be driven into the CO clouds.
Comparison of the metal abundances with the average stellar yields predicted by 
theoretical models of SN explosions points to SNe of Type II as the main
contributors of metals to the hot ISM.
There is no evidence of any correlation between radio-optical star-formation 
indicators and the measured metal abundances. Although due to uncertainties 
in the average 
gas density we cannot exclude that mixing may have played an important role,
the short time required to produce the observed metal masses ($\la2$ Myr) 
suggests that the correlations are unlikely to have been destroyed by
efficient mixing. More likely, a significant fraction of SN~II ejecta may be in a 
cool phase, in grains, or escaping in hot winds. In each case, any such fraction of
the ejecta would remain undetectable with soft X-ray observations.
\end{abstract}



\section{Introduction}

Massive stars deeply influence the baryonic component
of the Universe. Their ionizing radiation and
the supernovae (SNe) return kinetic energy and metal-enriched gas to the
interstellar medium (ISM) from which the stars formed (a process
usually called ``feedback''). 
Feedback exercises an influence not only on the gas-phase conditions in the immediate
environment of the clusters hosting those massive stars 
(e.g. McKee 1995; Wiseman \& Ho 1998; Pudritz \& Fiege 2000), 
but also on the phase structure and energetics of the ISM on galactic scales
(e.g. McKee \& Ostriker 1977; Norman \& Ferrara 1996)
and on the thermodynamics and enrichment of the inter-galactic medium (IGM)
on scales of several Mpc (e.g. Chiang, Ryu, \& Vishniac 1988; 
Heckman 1999; Aguirre et al. 2001).

The vast range of spatial scales involved is only one of the various difficulties 
encountered in attempting to study feedback. Even when one restricts the discussion
to purely mechanical feedback from SNe and stellar winds (commonly 
termed SN feedback), another difficulty is the broad range of
complicated gas-phase physics, which includes (magneto)hydrodynamic
effects such as shocks and turbulence, thermal conduction, and non-ionization
equilibrium (NEI) emission processes. 
Yet another complication is that much of the energy and
metal-enriched material involved is in the hard-to-observe coronal 
($T \ga 10^{5}$ K) and hot ($T \ga 10^{6}$ K) gas phases, requiring the
use of space-based far ultra-violet (FUV) and X-ray telescopes.

Several attempts at putting quantitative constraints on the metal enrichment
caused by SN feedback in starburst galaxies have been made in the recent
past. 
From a small sample of edge-on starburst galaxies observed by ROSAT
and ASCA, Weaver, Heckman, \& Dahlem (2000) obtained abundances completely 
consistent with solar values and no evidence of super-solar ratios of the
$\alpha$ elements with respect to Fe, as required by type-II SN feedback
(e.g., Gibson, Lowenstein, \& Mushotzky 1997). Moreover, they concluded that
the technique of measuring abundances through X-ray spectral fitting is
highly uncertain because of ambiguities in the fits (e.g., a degeneracy 
between the temperature and metallicity), and strongly dependent on the
model choice (see also Dahlem, Weaver, \& Heckman 1998; Dahlem et al. 2000;
Strickland et al. 2002, 2004).
The advent of Chandra (Weisskopf, O' Dell \& van Speybroeck
1996), with its sub-arcsecond angular resolution, allowed a better
subtraction of the point sources from the galactic diffuse emission, simplifying
the spectral fitting.  However, the low resolution spectra of the Chandra ACIS CCDs 
(Weisskopf et al. 1996) did not
allow a dramatic improvement in the accuracy of abundance measures. 
Notwithstanding the limits of ACIS spectral resolution, 
Martin, Kobulnicky, \& Heckman (2002), observing the dwarf starburst galaxy NGC~1569,
claimed to obtain ratios of $\alpha$ elements to Fe 2\,--\,4 times higher than the
solar value. They could not put constraints on the individual $\alpha$ element
abundances because of degeneracies between metallicity and spectral normalization.
Super-solar $\alpha$ elements to Fe ratios and super-solar abundances of individual $\alpha$ 
elements (e.g., Mg, Si and S) were also reported in other galaxies
observed by Chandra, e.g. the face-on spirals NGC~6946 (Schlegel, Holt, \& Petre 2003)
and M~83 (Soria \& Wu 2003).

While X-ray observations of the halos of nearby edge-on disk galaxies
may provide the best single probe of the action of mechanical feedback 
on galactic scales, observations of the disks of {\it face-on} nearby spiral 
galaxies may provide useful insights into the physics and enrichment of the hot gas.
Such observations allow a spatially resolved analysis of the hot ISM in the disk
that is much less affected by internal absorption than in the case of edge-on galaxies.

The Antennae are the nearest pair of colliding, relatively face-on spiral galaxies
involved in a major merger (D = 19~Mpc for $H_0 = 75~\rm km~s^{-1}~Mpc^{-1}$).
Hence, this system not only allows a study of the properties of the ISM relatively
unaffected by internal absorption, but it also provides a unique 
opportunity to get the most detailed insight possible into the consequences 
of a galaxy merger, such as induced star formation and its effect on the ISM.

The presence of an abundant hot ISM in The Antennae was originally suggested by
the first {\it Einstein} observations of this system (Fabbiano \&
Trinchieri 1983), and has since been confirmed by observations with several
major X-ray telescopes ({\it ROSAT}: 
Read, Ponman, \& Wolstencroft 1995, Fabbiano, Schweizer, \& Mackie 1997; and 
{\it ASCA}: Sansom et al.\ 1996).
The first {\it Chandra} ACIS  observation of The Antennae in 1999 gave us for the first time a detailed 
look at this hot ISM, revealing a complex, 
diffuse, and soft emission component responsible for about half of the 
detected X-ray photons from the two merging galaxies (Fabbiano, Zezas \& Murray 2001;
Fabbiano et al.\ 2003, hereafter F03).
The spatial resolution of {\it Chandra} is at least
10 times superior to that of any previous X-ray observatory, which allows us to 
resolve the emission on physical scales of $\sim$75~pc (for D$=$19~Mpc) and to 
detect and subtract individual point-like sources (most likely X-ray binaries; 
see Zezas et al.\ 2002).
At the same time, the {\it spectral} resolution of ACIS also allows us to study for 
the first time the X-ray spectral properties of the various
emission regions, providing important additional constraints on their nature.

While the first {\it Chandra} data set demonstrated the richness of the ISM
in The Antennae, the number of detected photons was insufficient to study its
detailed small-scale morphology and spectral properties.
A deep monitoring observing campaign of The Antennae with
{\it Chandra} ACIS has produced a detailed and rich data set.
A first look at this data-set revealed a complex diffuse emission,
with signatures of strong line emission coming from some of the hot ISM regions, 
pointing to high $\alpha$ element abundances (Fabbiano et al.\ 2004). 
This paper focusses on the physical properties of the hot gas and its
enrichment by SN explosions.
In a companion paper we describe the data-analysis procedure, the 
creation of broad-band mapped-color images of the diffuse emission, and the 
performed spectral analysis (Baldi et al. submitted, hereafter Paper I), 

This paper is organized as follows.
In \S2 we briefly review the main results of Paper I.
In \S3 we derive the hot gas parameters, like pressure and density, from the spectral 
fits.
In \S4 we discuss the main results derived from the 
abundance measures, focusing especially on the supernova feedback. 
In \S5 we present some further insights into the physical properties of the hot ISM,
and in \S6 we summarize the main results.

\section{Mapped-color images of the diffuse emission and main results of the spectral analysis}

NGC 4038/39 was observed with {\it Chandra} ACIS-S seven times during the period 
between December 1999 and November 2002, for a total of $\sim$411 ks. 
The data products were analyzed with the CXC CIAO v3.0.1 software and 
XSPEC package.
Full details about the data processing and reduction are given in Paper I.

Paper I presents a mapped-color image of the diffuse emission in The Antennae,
representing contributions from three different energy bands: 
0.3--0.65 keV, 0.65--1.5 keV and 1.5--6 keV.
Before combining the images from the three bands,
all point sources were subtracted and an adaptive gaussian kernel was applied to
each of the three images.
Figure~\ref{truecolor} reproduces the mapped-color image for the reader's convenience.
Red denotes soft X-ray emission (0.3--0.65 keV), green intermediate-energy emission
(0.65--1.5 keV), and blue hard X-ray emission (1.5--6 keV).
A similar procedure was applied to generate a line-strength map for the
emission lines from O+Fe+Ne, Mg, and Si in different regions of the hot ISM.
Figure~\ref{linemap} shows this second map, where red, green, and blue represent
O+Fe+Ne emission (0.6--1.16~keV), Mg emission (1.27--1.38~keV), and Si emission 
(1.75--1.95~keV), respectively.

Both the mapped-color image of the diffuse emission and the line-strength 
map guided us in selecting 21 spectrally similar regions
for a proper spectral analysis.  These regions are shown in Figure~\ref{regions}.
The method used for the spectral extraction and analysis is fully described in 
Paper I. Table~\ref{mytable1} resumes the main results of the spectral analysis
conducted so far.
In this table, the goodness of fit ($\chi^2_{red}$ over the number of degrees of freedom), 
absorbing column density $N_H$, temperature(s) $kT$, abundances of Neon, Magnesium, 
Silicon, and Iron, and power-law photon index $\Gamma$ (if a power-law component 
is necessary) are summarized together with the ``best-fit model'' 
(as defined in \S5.2 of Paper I) for each region. 

Almost all regions could be fitted with a single thermal component.
Only Region 4b required two thermal components with different $kT$;
in this region, we could not constrain the low $kT$ value (although we found
a minimum in the $\chi^2$ statistics at $kT=0.20$ keV), and we could put only a lower
limit ($kT>0.54$ keV) to the higher temperature component. 
Almost all the single-temperature regions required a temperature of $kT\sim0.6$ keV
to fit the data, except for regions 5, 8b, and 14 ($kT\sim0.3$ keV).
Considering all the regions analyzed, the temperatures (both in the single- and 
two-temperature cases) range from $\sim0.2$ to $\sim0.8$ keV (including
the errors).
Our analysis yielded metal abundances that have acceptable constraints in the 
majority of regions, and led to the most important, and somewhat unexpected, 
result of our spectral
analysis: the abundances of Silicon and Magnesium cover a wide range of values 
ranging from the very low values found in regions 2, 6a, 12a, and 12b 
($Z\sim 0.2 Z_\odot$) to the very high values observed in Region 5 ($Z\ga 20 Z_\odot$).
The intrinsic absorption throughout the hot ISM is generally low, with 
$N_H\sim10^{20}$ cm$^{-2}$ typical and often consistent with zero.
The exceptions are the southern nucleus (regions 8a and 8b) and Region 7,
which is significantly obscured also in the optical.
In the optical and near infrared, Region 7 corresponds to
the ``Overlap Region,''
where the most active star formation is now occurring (Mirabel et al.\ 1998; 
Wilson et al.\ 2000; Zhang, Fall, \& Whitmore 2001). These three regions show
intrinsic absorption of $\sim1-2\times10^{21}$ cm$^{-2}$.

\section{Hot-gas parameters}\label{hotparam}

All the calculations performed in this section use the parameters obtained from the 
best-fit spectral models and listed in Table~\ref{mytable1}.\\

Table~\ref{lumin} gives the detailed emission parameters derived from the spectral 
analysis of each region.
The observed flux in the 0.3--10 keV band is listed for both the power-law (if present) 
and the thermal emission in columns (2) and (3), respectively.
The emission measure ($EM=n^2V$) is listed in column (4). 
For each region we present also the intrinsic (absorption-corrected for $N_H$) 
luminosity in the 0.3--10 keV band for both the power-law (if present) 
and the thermal emission (columns (5) and (6)).
Column (7) gives the 
ratio between the intrinsic luminosities in the 0.3--10 keV band of the power-law and 
thermal components.
In the case of Region 4b, modelled 
with two temperatures, we were unable to determine a confidence range for $EM$
because of the lack of constraints on both temperatures ($kT_1$ 
unconstrained and $kT_2>0.54$ keV at 1$\sigma$). Therefore, we give only the values
corresponding to the best-fit $kT_1$ and $kT_2$ (corresponding to 0.20 keV and 0.60 keV,
respectively)\\

>From these parameters and those listed in Table~\ref{mytable1}, we can estimate
various properties of the
emitting plasma, such as electron density $n_e$ ($\approx n_H$), 
thermal energy content $E_{th}$, cooling 
time $\tau_c$, and pressure
$p$. These quantities are given in Table~\ref{mytable2}. The calculation of 
these parameters depends on the emitting volume and on the filling factor
$\eta$ of the hot gas. Since we observe
regions projected on the plane of the sky, we can only measure the ``footprint''
of each emitting region.
To estimate the emitting volumes we adopted the following procedure. For
all ``single''-temperature regions, we assumed a depth of $200K$ pc for the 
emitting volume, where $K$ is a variable parameter (the numbers given in 
Table~\ref{mytable2} assume $K=1$, corresponding to the 
typical depth of a spiral disk).
In the case of Region 4b, where we have two temperatures, we assumed that the
high- and low-temperature components are in 
pressure equilibrium, and derived the thicknesses for the volumes occupied by 
these two phases from the best-fit $kT$s, imposing the condition that
they add up to a total value of $200K$ pc.
The physical area covered by each region scaled to the distance of The
Antennae is given in column (2). 
Column (3) lists the thermal gas electron
density $n_e$. This quantity depends both on the depth-variable parameter
$K$ and on the filling factor $\eta$ of the gas and is proportional to 
$\eta^{-1/2}K^{-1/2}$.
The thermal energy $E_{th}$ and the cooling time $\tau_c$ (columns (4) and (5), respectively)
were estimated following Tucker (1975). Both these quantities are proportional 
to $\eta^{1/2}K^{1/2}$.
The thermal pressure of the hot gas, expressed in dyne cm$^{-2}$, is listed 
in column (6).
The pressure ($p=2nkT$), being directly proportional to the density $n_e$, has 
the same dependence
on the variable-depth parameter $K$ and filling factor $\eta$ as the 
density ($\eta^{-1/2}K^{-1/2}$).
The filling factor $\eta$ is assumed to be unity in our calculations.
A filling factor $\eta\la1$ for the hot phase of the ISM is plausible 
from recent results coming from 3D hydrodynamical simulations ($\eta$ ranging
between 0.17 and 0.44 for SN rate between Galactic and 16 times the Galactic
value; see, e.g., de Avillez \& Breitschwerdt 2004 and references therein). 
However, our parameters have only a slight dependence on it ($\propto\eta^{-1/2}$ or
$\propto\eta^{1/2}$), hence our estimates are precise within a factor of two or 
less.
As described above for the emission measure $EM$, in Region 4b we were unable to compute
a confidence range for
$n_e$, $E_{th}$, $\tau_c$, and $p$ because of the indeterminacy of the two
temperatures. The values of these parameters for Region 4b correspond
to the best-fit values of $kT_1$ and $kT_2$.
Metz et al. (2004) found somewhat larger thermal energies
for regions similar to ours.  This probably results from
their smaller density estimates.\\

The calculated cooling times, in the $10^7$--$10^8$ yr range, are in agreement with the 
results of F03.
The pressure is generally of the order of a few $10^{-11}$
dyne cm$^{-2}$, but reaches higher values (a few $10^{-10}$ dyne cm$^{-2}$) 
in the two nuclei and also in the regions corresponding to the hot-spot regions
R1 and R2 of F03. These new values are in excess of those found in F03,
possibly reflecting the fact that larger areas were averaged over in that work.
To test this hypothesis we extracted a spectrum for each of these regions using the 
same region files as in F03.
The pressures we obtain for all four regions are, indeed, lower than our new estimates in 
the corresponding regions,
although still higher than the pressure estimates in F03.
F03 argued for a 
possible pressure equilibrium between the hot gas and the cold molecular clouds.
Zhu, Seaquist, \& Kuno (2003) derived from CO measurements a pressure of 
$4.2\times10^{-11}$ for the northern nucleus (NGC 4038) and of $3.1\times10^{-11}$ 
dyne cm$^{-2}$ for the southern nucleus (NGC 4039).
Our observations suggest that the hot-gas pressure is a few times higher than the CO estimate for
the northern nucleus and almost an order of magnitude higher in the southern nucleus.
Such a large pressure difference would imply that shock waves are being driven 
into the CO clouds. The hot ISM may therefore in part be responsible for compressing 
and fragmenting the CO clouds, triggering star formation.
However, we must bear in mind
the large uncertainties in the estimated pressures, due to the assumptions we had to make 
on the emitting volume and the small-scale properties of the clouds.\\

Table~\ref{masses} presents our estimates of the total mass of hot gas and of the masses
of the individual elements Ne, Mg, Si, and Fe for each region.
The calculation of these individual-element masses was performed using 
the Anders \& Grevesse (1989) compilation of meteoritic abundances.
All values in Table~\ref{masses} are dependent on both $K$ and $\eta$ and are 
proportional to $\eta^{1/2}K^{1/2}$.
For both the northern and the southern nucleus our total mass estimates are a 
factor of two smaller than the values previously obtained in F03. This discrepancy may be 
due to differences in the best-fit models
and to the sizes of the regions used for the spectrum extraction, which were 
larger in the F03
analysis.
The total mass of hot gas derived by adding the contributions from all individual regions is 
$\sim1.3\times10^7$
M$_\odot$. For the individual elements we obtain total masses of
$\sim3.4\times10^4$ M$_\odot$ for Ne, $\sim1.3\times10^4$ M$_\odot$ for Mg,
$\sim2.6\times10^4$ M$_\odot$ for Si, and $\sim1.8\times10^4$ M$_\odot$ for Fe.

It is worthwile to note that, if we adopt the smaller distance of $\sim$14 Mpc for The Antennae,
as derived from the tip of the red giant branch by Saviane, Hibbard, \& Rich (2004), some of the
above physical parameters would change.
In particular, thermal energies, luminosities, and masses would be smaller by a factor
of almost 2, while all other listed quantities would remain unvaried.

\section{Abundances}\label{sectabund}

The unprecedented high angular resolution and sensitivity of {\it Chandra}, in 
conjunction with the long integrated-exposure time of our observations, allows 
us to analyze in detail the properties of the hot ISM in The Antennae, down to 
regions measuring a few hundred parsecs in size.  Based on only the first 75~ksec observation 
of 1999 December, F03 were unable to constrain the variety of metal abundances in these regions. 
Indeed, the measure of hot gas elemental abundances in star forming galaxies 
has always been a challenging task (e.g., Strickland \& Stevens 1998, 2000;
Dahlem, Weaver \& Heckman 1998; Weaver et al. 2000; Dahlem et al. 2000;
Strickland et al. 2002; 2004). 
However, the new, deeper total exposure has allowed us, within
the limitations of the low spectral resolution of the ACIS CCDs, to place better
constraints on the abundances than in the past. The method we have used to determine the abundances
and their relative confidence ranges, described in Paper I, 
is conservative enough to take into account both statistical and systematic biases 
that may affect the abundance measures in our low-resolution spectra, especially regarding the
biases deriving from model choice and from the metallicity-temperature degeneracy.
In the past, spatially resolved measurements of individual elemental abundances 
were performed on megaparsec-size regions in the hot intra-cluster medium of 
galaxy clusters (e.g., Mushotzky et al.\ 1996; Finoguenov, David, \& Ponman 2000)
and, on a much smaller scale, in elliptical galaxies (e.g. Finoguenov et al. 2002). 
With the new {\it Chandra} data for The Antennae, we are now trying to explore the ``local''
chemical enrichment of the hot ISM by supernovae within two interacting galaxies.\\

As discussed in Paper I, the elemental abundances of Ne-IX, Mg-XI, Si-XIII, and
Fe-L that we measure in the hot ISM of The Antennae are generally consistent with
the stellar abundances measured from optical data
($\sim$0.5 solar; Fritze-v.Alvensleben 1998), except for a few regions (4b, 5, 7, 8a and 8b)
where the measured abundances are clearly in excess of this value and, in the case of Region 5,
even significantly super-solar.
Indeed difficulties in measuring the abundances through X-ray spectral fitting of low 
resolution spectra are well documented in the literature (e.g., Weaver
et al.\ 2000). Our very conservative approach, however, did not prevent us from locating
peculiar regions with bona-fide super-solar abundances. While all the other regions have
metallicities consistent with solar (within the errors), Region 5 is showing particularly
enhanced abundances for three out of the four elements we measured ($Z_{Ne}>3.9$, $Z_{Mg}>3.8$,
$Z_{Fe}>1.3$, at 90\% confidence level). Such high X-ray derived metallicities of Fe and 
especially of the $\alpha$ elements are quite rare and have not often been found in previous
observations of other galaxies. This could be
due both to the peculiarity of Region 5 and to the fact that 
constraining abundances through X-ray fitting of low resolution spectra
is always a challenging task. 
From a Chandra observation of the face-on spiral NGC~6946 Schlegel et al.\ (2003) 
found an abundance of Si$=3.00_{-1.90}^{+1.95}$ for a single-temperature model fit
of the diffuse emission in this galaxy. Although the errors they quote are
at 90\%, they estimated this confidence level considering only two interesting parameters,
a method a lot less conservative than ours. Hints of super-solar values of the $\alpha$
elements abundances were also found in the starburst spiral galaxy M~83 by Soria
\& Wu (2002). However, these authors did not fit the abundances, but simply realized that 
a single-temperature thermal model with the $\alpha$ elements abundances fixed at
super-solar values and an $\alpha$ to Fe ratio higher than solar (e.g., $Z_{Ne}=2.0$,
$Z_{Mg}=2.5$, $Z_{Si}=3.0$, $Z_{Fe}=0.9$) gives clearly a better fit ($\chi^2_{red}\sim0.99$)
than a model with all the elements fixed at solar ($\chi^2_{red}\sim1.42$).
All the other measurements derived from Chandra ACIS observations of starburst galaxies
or ultra-luminous infra-red galaxies (ULIRGs) and presented recently
in the literature led to abundances consistent with solar (e.g., the ULIRG sample of Huo et
al. 2004; the edge-on starbursts sample of Strickland et al. 2004) or clearly 
subsolar (e.g. NGC~253; Strickland et al. 2002).
\\

\begin{figure}[h]
\plottwo{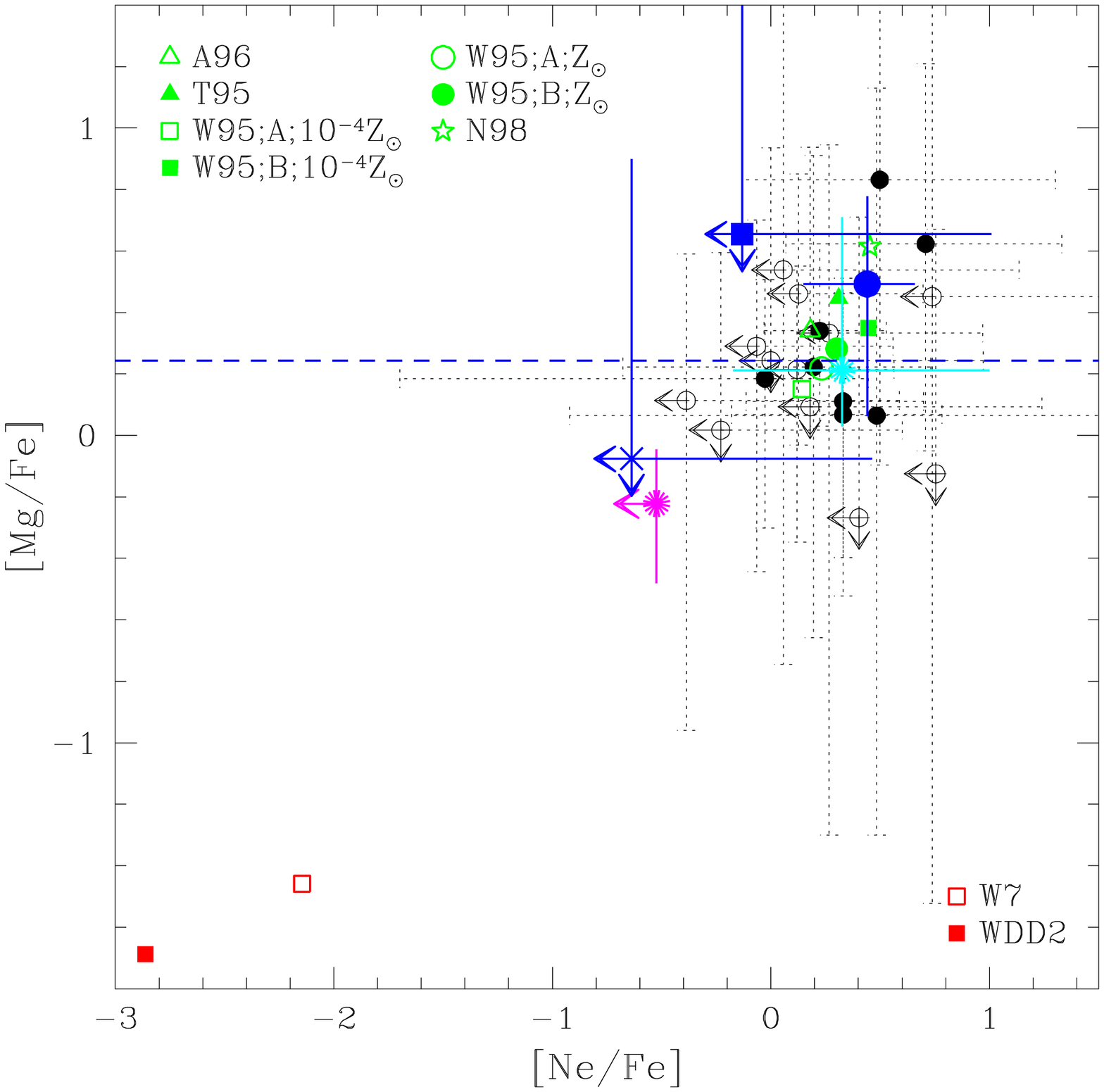}{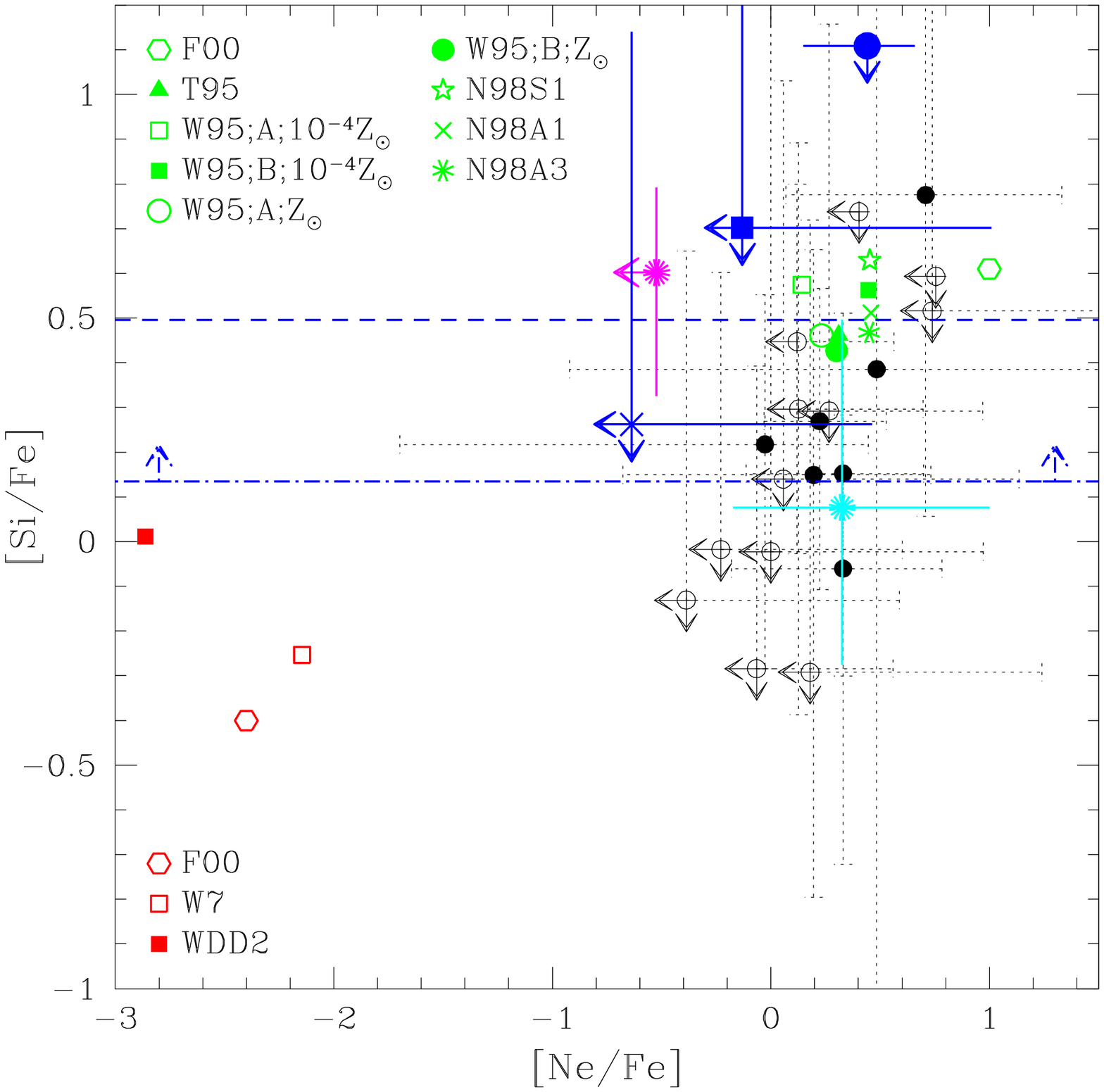}
\epsscale{0.40}
\plotone{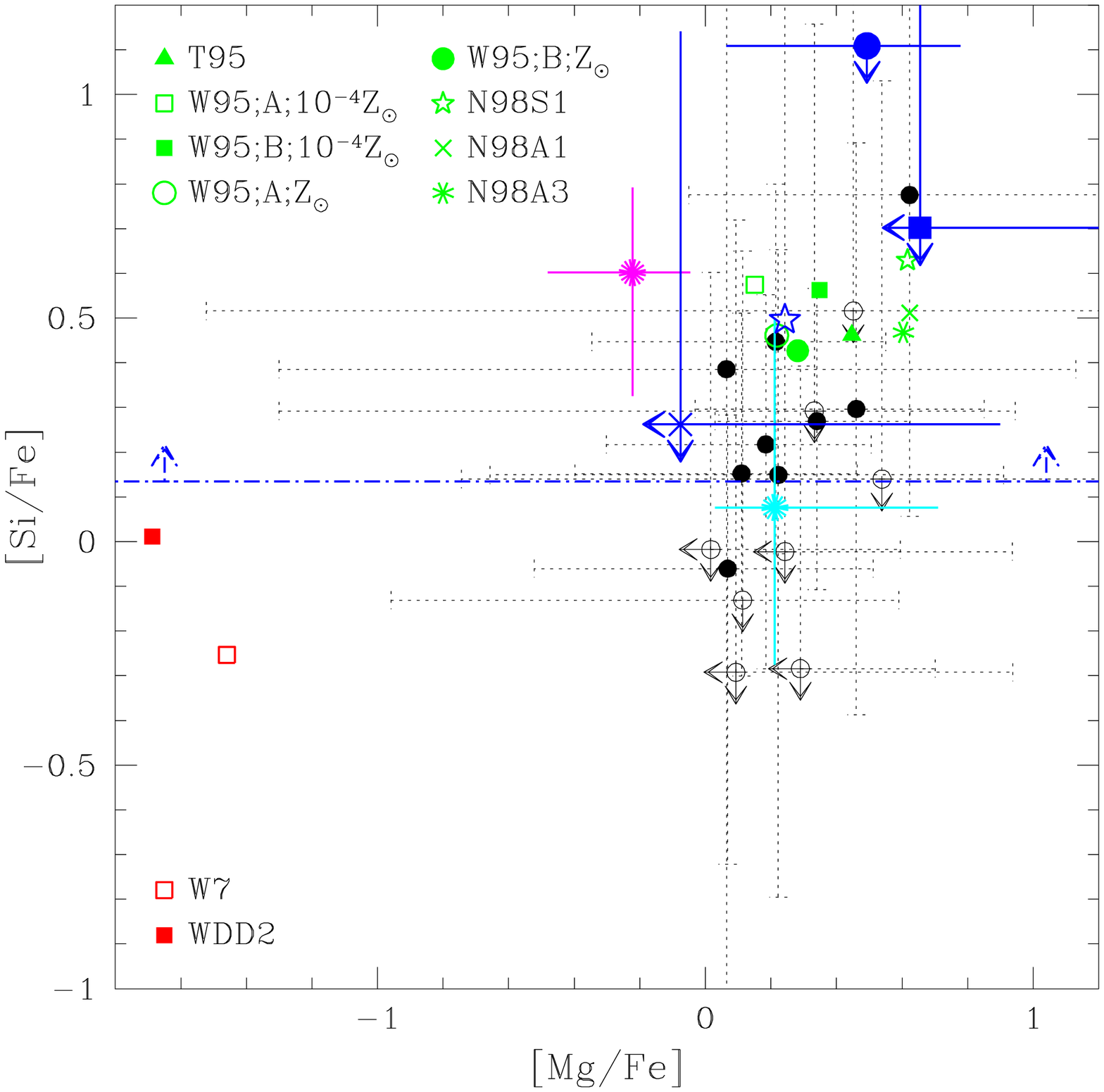}
\caption{Abundance-ratio diagrams for regions of the 
Antennae hot ISM: (a) [Ne/Fe] vs.\ [Mg/Fe]; (b) [Ne/Fe] vs.\ [Si/Fe];
(c) [Mg/Fe] vs.\ [Si/Fe]. 
Abundance units refer to meteoritic abundances of Anders \& Grevesse (1989). 
The errors are at $68\%$, and brackets represent the logarithmic values. Each 
one of the regions in Table~\ref{mytable1} is plotted with a black dot, empty 
if the detection significance of the Mg or Si line in the X-ray 
spectrum is less than 2$\sigma$.  Upper limits are plotted with black 
arrows.
Red symbols correspond to ratios expected if all the elements were generated 
exclusively by SNe~Ia, while green symbols refer to the average stellar 
yield of SNe~II. ``F00'' indicates values adopted in Finoguenov et al.\ 
(2000), while the other
labels refer to the theoretical work listed in Nagataki \& Sato (1998).
The magenta and cyan symbols indicate abundance ratios for Cas A 
and N132D, derived from
the metallicities observed by BeppoSAX (Favata et al.\ 1997a, 1997b).
Blue symbols and lines refer to determinations of abundance ratios in 
other Galaxies:
the warm Galactic halo (star symbol or dashed line), NGC~253 (filled square
for the northern halo, cross for the disk emission), NGC~6946 (lower limit
on [Si/Fe] represented as a dash-dotted line) and the Strickland et al. (2004) 
sample of edge-on starbursts (filled
circle).
\label{SNyields}}
\end{figure}

The ratios between elemental abundances in each region can be related to the
type of supernovae generating the metals.
If the elemental enrichment results from Type II supernovae, one
would expect [Ne/Fe] and [Mg/Fe] values approaching 0.3, and [Si/Fe] values approaching 0.5 
on average; the corresponding values expected in Type Ia supernova enrichment are
dramatically lower (Nagataki \& Sato 1998). We used our results to explore
the origin of the metals in The Antennae.
We calculated the [Ne/Fe], [Mg/Fe], and [Si/Fe] ratios of the regions listed 
in Table~\ref{mytable1} and plotted them
in diagrams displaying the [Ne/Fe] ratio vs.\ [Mg/Fe] ratio (Figure~\ref{SNyields}a), 
[Ne/Fe] vs.\ [Si/Fe] (Figure~\ref{SNyields}b), and [Mg/Fe] vs.\ [Si/Fe] 
(Figure~\ref{SNyields}c).
In these diagrams, we also show the theoretical stellar yields expected
from SNe~Ia (red symbols) and SNe~II (green symbols), taken from the compilations 
by Finoguenov et al.\ (2000) and Nagataki \& Sato (1998).
We also indicate the regions occupied by the type-II supernova 
remnants (SNR) Cas A and N132D, using the abundances measured by Favata et al.\ (1997a,b) 
from BeppoSAX data.
To relate our results to other measurements of these ratios in the literature we
plot also the values relative to the warm Galactic halo (Savage \& Sembach 1996),
to the emission both from the disk and from the northern halo of NGC~253 (Strickland et al.\ 2002),
to the X-ray diffuse emission observed in NGC~6946 (Schlegel et al.\ 2003), and to the
averaged value in the hot halo of a sample of ten nearby edge-on starbursts (Strickland et
al.\ 2004).

Although the uncertainties are often large, almost all regions observed in The Antennae
show [Ne/Fe] and [Mg/Fe] ratios fully consistent with a SN~II-dominated 
enrichment scenario.
However, the [Si/Fe] ratio is clearly lower than expected in a SN~II enrichment 
scenario,
and in some cases it could be consistent with SN~Ia enrichment, although the 
uncertainties are quite
large also for this ratio.  The only two exceptions to these trends are 
regions 5 and 7, both
located in the upper right corners of the diagrams.  These two regions
have a [Si/Fe] ratio fully consistent with an enrichment
stemming entirely from type-II supernovae. Region 7 is the ``Overlap Region'' of the
two merging galaxies, where the most active star formation
is presently occurring (Mirabel et al.\ 1998; Wilson et al.\ 2000; Zhang et al.\ 
2001).  It seems plausible that recent episodes of star formation
led to the injection of hot enriched gas into the ISM almost exclusively 
through SNe~II.
Region 5 is the region with the highest abundances of both Mg and Si.
Similarly to Region 7, Region 5 has a high relative star-formation strength and star-formation
rate per unit 
area (see Table~\ref{francois2} and discussion below).

There are five regions with very low [Si/Fe] values located 
in the lower part of Figures~\ref{SNyields}b and c
([Si/Fe] $<0$, even encompassing the confidence range).
These low [Si/Fe] values appear to be 
inconsistent with SN~II enrichment and more typical of SN~Ia enrichment.
We caution that this surprising result may be affected by poor statistics.  Typically,
in these regions the Si line is detected at less than 2$\sigma$ significance 
level, in some cases because of the poor statistics at $E\ga 2$ keV that may 
lead to an underestimation of the Si abundance. However note that this deficiency
may be an independent confirmation of the failure of type II SN yields
predictions to fit the spectra (e.g. linking together all the $\alpha$ elements 
using the type II SNe ratio from the compilation of average stellar yields from
Nagataki \& Sato (1998); see Paper I).


While there may be some uncertainties both in the SN yield models and (even
more) in our abundance measures,
we may also be seeing the effect of different mean
ages of star-formation regions and, therefore, of different distributions of
SN masses in different X-ray emission regions. 
This observed Si deficiency may also be explained if a significant fraction of the Si, larger
than for the other elements, has cooled and is locked in dust grains.
Silicon is one of the most refractory elements, while Neon
is one of the least refractory elements, as indicated by depletions
derived from interstellar absorption lines
(see Savage \& Sembach 1996 for a review of this topic). 
However, Iron is depleted at least as strongly as Silicon,
and there is no supporting observational or theoretical evidence to claim that 
Silicon should be more strongly affected than Fe by cooling and locking up in dust.

We note that the abundance ratios measured in the majority of 
the NGC 4038/39 regions are consistent with ratios observed in SNR N132D. On the other hand, 
the measured ratios do not agree with the values observed in Cas A. The 
different measured ratios in 
these two SNRs, both originating from a type-II SN event, can give us an idea of
the complexity and wide variety of input physics involved in
yield predictions (e.g., convection, reaction rates, method for 
initiating shock waves; see Gibson et al.\ 1997).

Although our measurements are affected by large uncertainties they represent clearly
an improvement over the previous efforts aimed at measuring the $\alpha$ to Fe
ratio in starburst galaxies. The only exception comes from the Strickland et al.\
(2004) sample of edge-on starburst galaxies. This data point, which has smaller 
error bars than our typical error bars, is consistent with the ratios observed in 
almost all the regions of The Antennae. However, note that their 
smaller error bars derive from 
the fact that they come from the observed hot halo emission in all ten galaxies of the
sample. Moreover, in this sample they cannot put a stringent upper limit on the [Si/Fe]
ratio, a thing that we were able to do in at least some of our regions.

\subsection{X-ray abundances vs. Optical/Radio SF indicators}

To further explore our results, we searched for correlations with age and star 
formation indicators for the different regions of The Antennae. 
NGC 4038/39 was observed by {\it HST}/WFPC2 (F336W, F439W, F555W, and F814W filters) 
in 1996 January (Whitmore et al.\ 1999). There were also
VLA observations (in configurations BnA, CnB, and B at 6 and 4 cm) performed 
in 1997 January and 1998 September (Neff \& Ulvestad 2000).
Based on the {\it HST}/WFPC2 data and VLA observations, we looked for 5 different 
types of objects within each of our regions, counting:
(1) the number $\#R$ of very red, extremely young clusters (Zhang et al.\ 2001,
84 in total),
(2) the number $\#B$ of bright young blue clusters (Table~1 in Whitmore et al.\
1999, 
50 in total),
(3) the number $\#I$ of intermediate age clusters (Table~2 in Whitmore et al.\
1999, 30 in total, of which 5 unpublished),
(4) the number $fl$ of flat-spectrum, thermal radio sources ($\alpha>-0.40$), 
and (5) the number $st$ of steep-spectrum, non-thermal radio sources 
($\alpha\leq-0.40$). In addition,
brightness estimates were performed visually on the {\it HST}/WFPC2 images, aimed at estimating
the integrated luminosity
of each region in the $U$, $V$, $I$, and $H\alpha$ passbands on a scale ranging from 
0 to 10.  

>From the published data and our visual estimates it was possible to derive---for each
region---four quantities that might correlate with ISM metal
abundances: a mean cluster age, the relative strength of star formation, the star
formation per unit area, and the number of SNRs per unit area.
The mean cluster age in each region was computed as the number-weighted 
average of the mean ages of very red clusters (4 Myr) and young blue clusters 
(15 Myr); we ignored the intermediate-age ($\sim$500 Myr) clusters 
because their small known numbers (0--2 per region) would have added excessive noise.
The relative strength of star formation was estimated via the ratio 
$(U+H\alpha)/(V+I)$, where
``relative'' is with respect to the older populations contributing light to the 
$V$ and $I$ passbands.
The star formation per unit area was calculated as the sum of $U$ and 
$H\alpha$ estimates
(on a scale of 0--10) divided by the estimated area of each region in kpc$^2$.
The number of SNRs per unit area was defined as the number 
of VLA steep-spectrum radio sources (from Neff \& Ulvestad 2000)
divided by the estimated area of the region.

Table~\ref{francois} summarizes our visual estimates, while Table~\ref{francois2} is a
summary of the derived quantities and also includes information
about the presence of major star-forming complexes named by Rubin, Ford, 
\& D'Odorico (1970) in its last two columns.

\begin{figure}[h]
\epsscale{0.90}
\plottwo{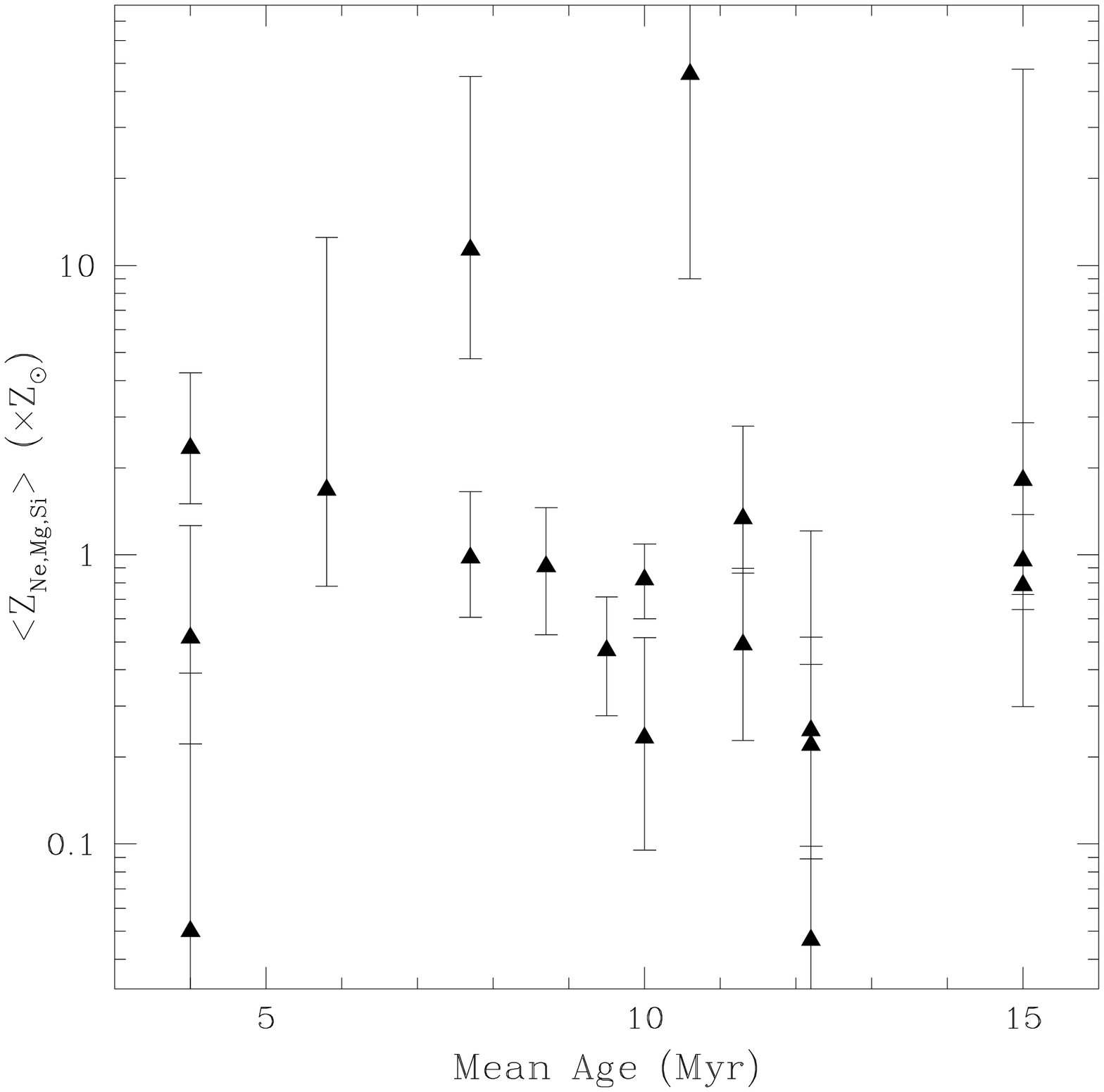}{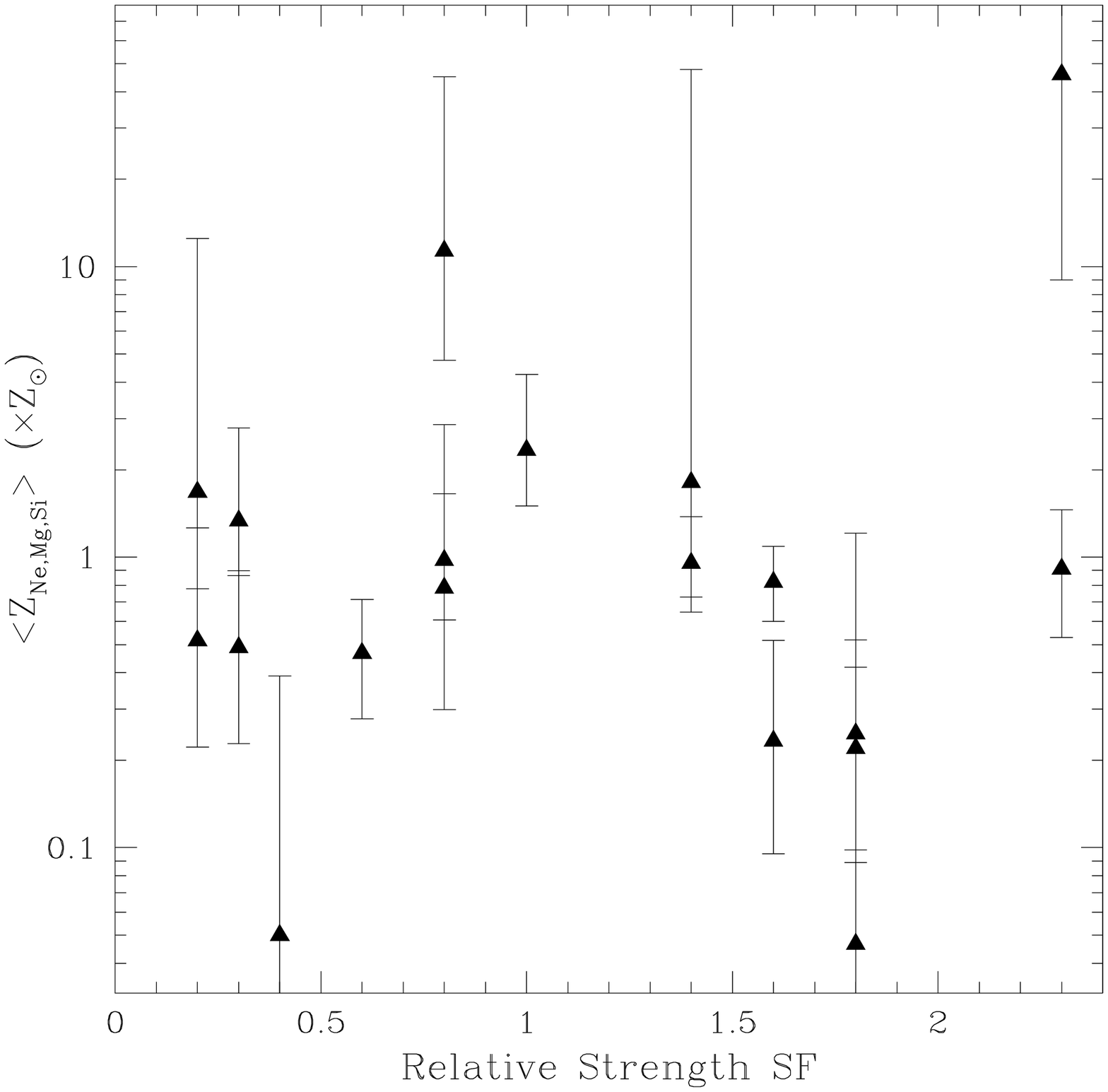}
\epsscale{0.90}
\plottwo{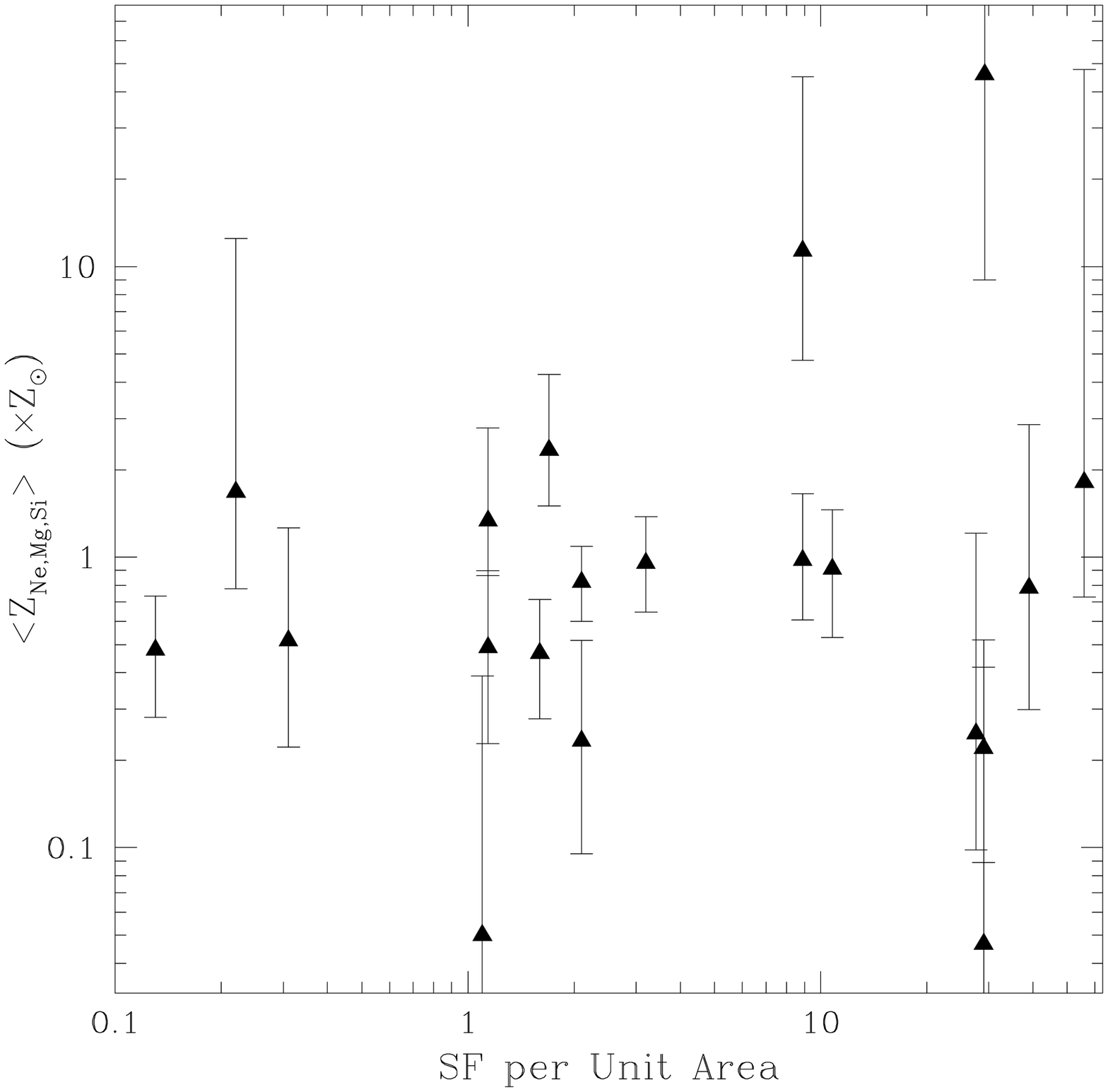}{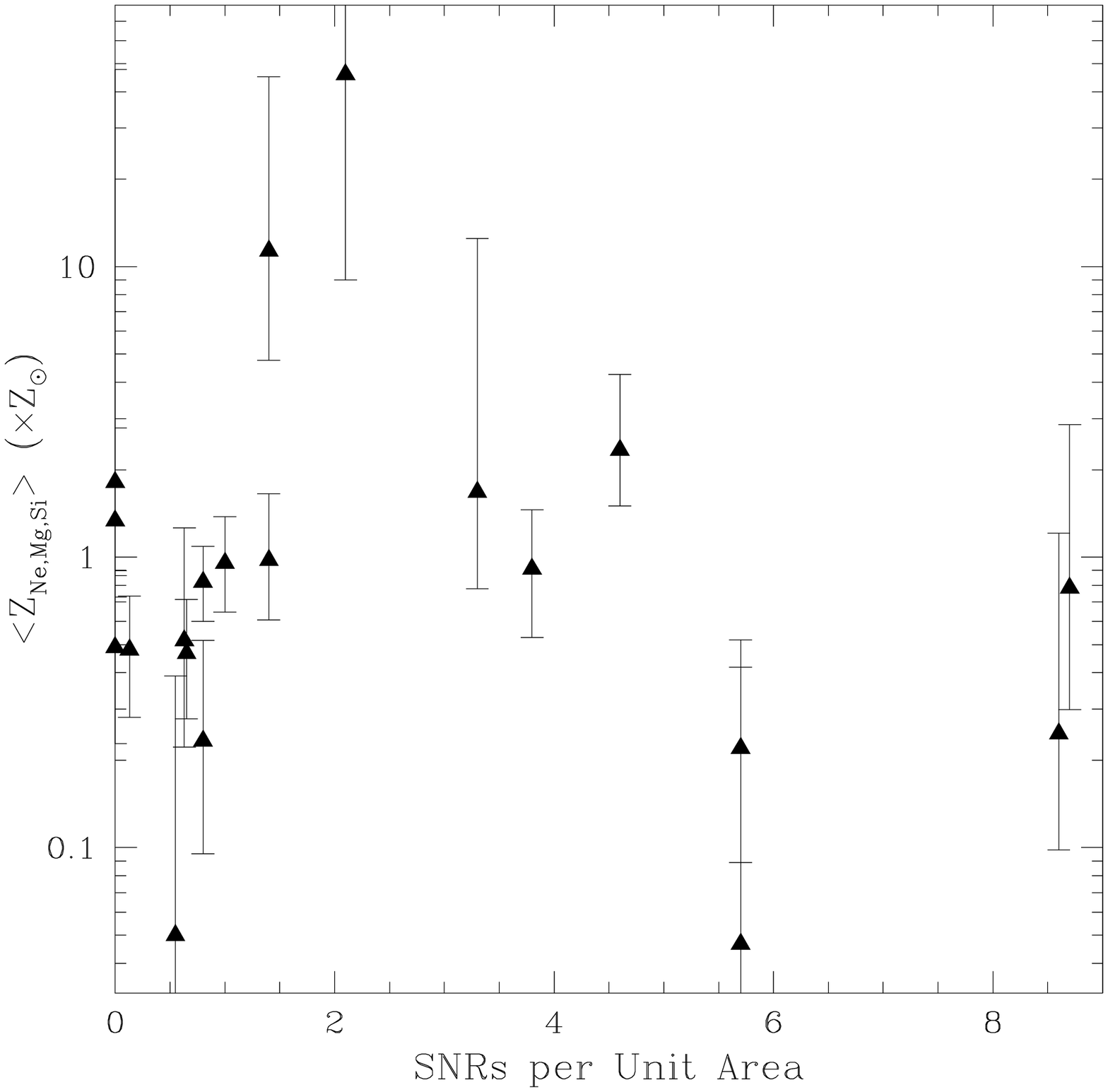}
\caption{Diagrams of the mean abundances of $\alpha$-elements in each region 
of the Antennae hot ISM plotted versus (a) mean age of the star clusters,
(b) relative strength of star formation, (c) star formation per unit area,
and (d) SNRs per unit area.
\label{optvsNeMgSi}}
\end{figure}

\begin{figure}[h]
\epsscale{0.90}
\plottwo{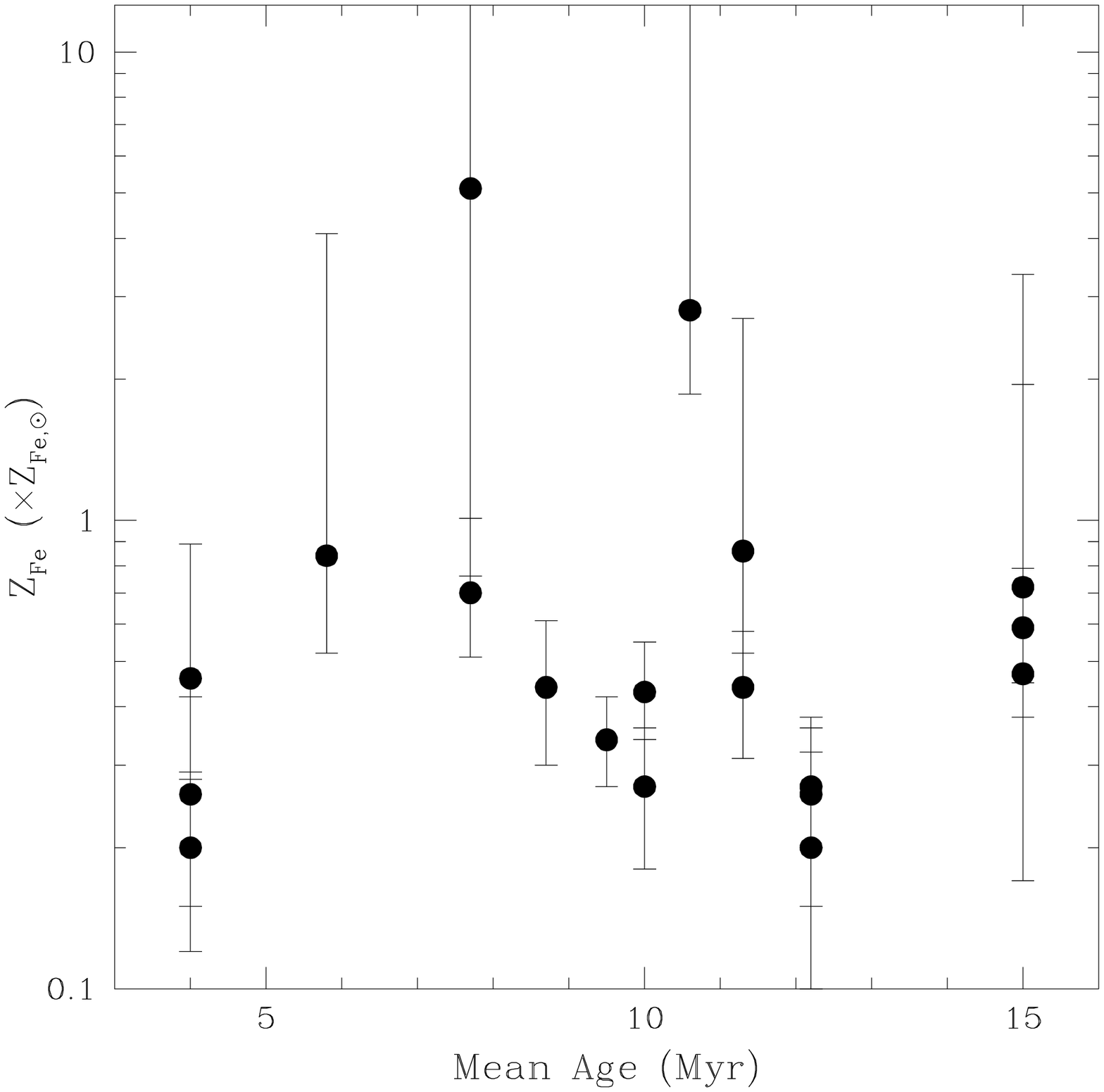}{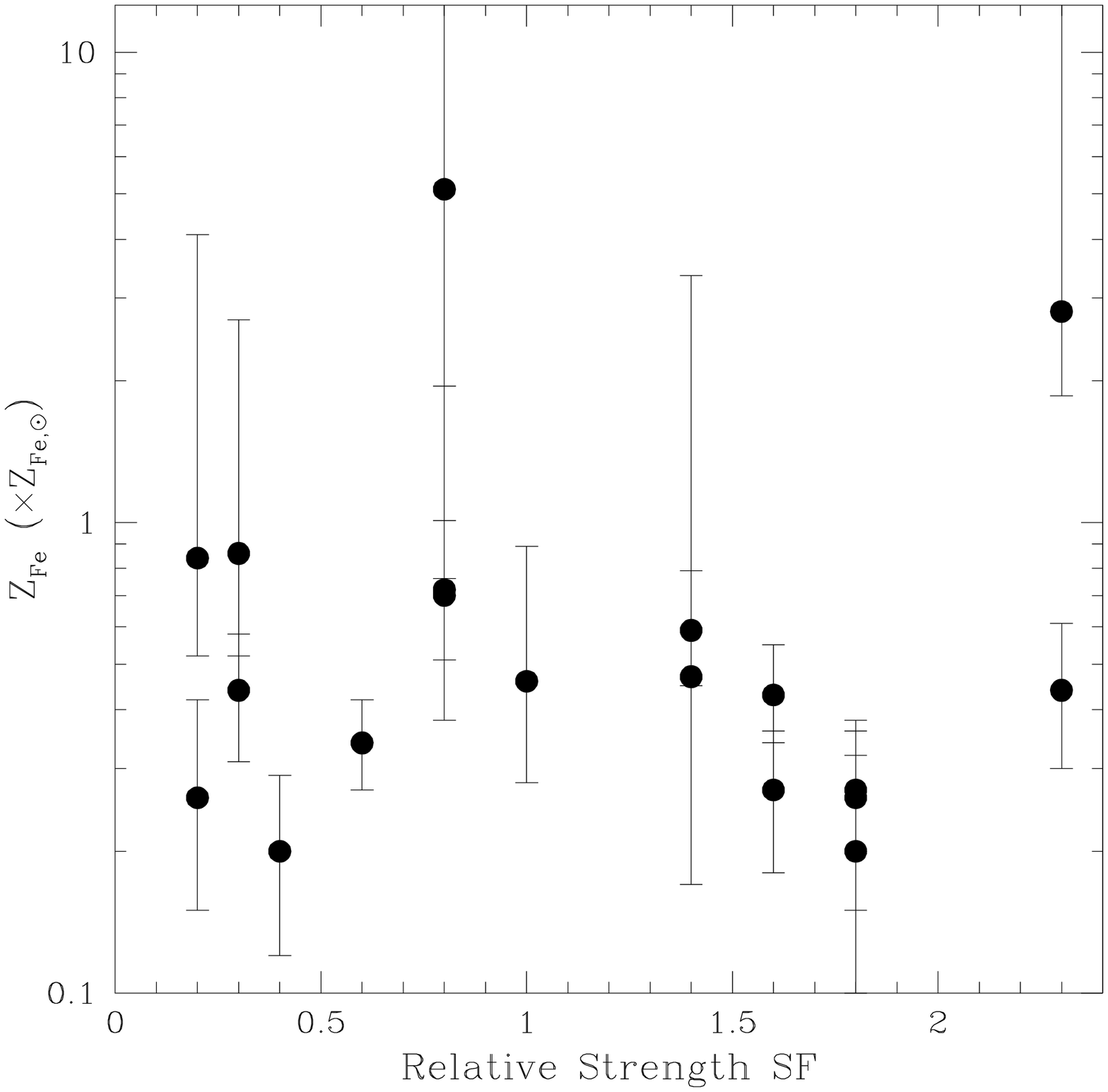}
\epsscale{0.90}
\plottwo{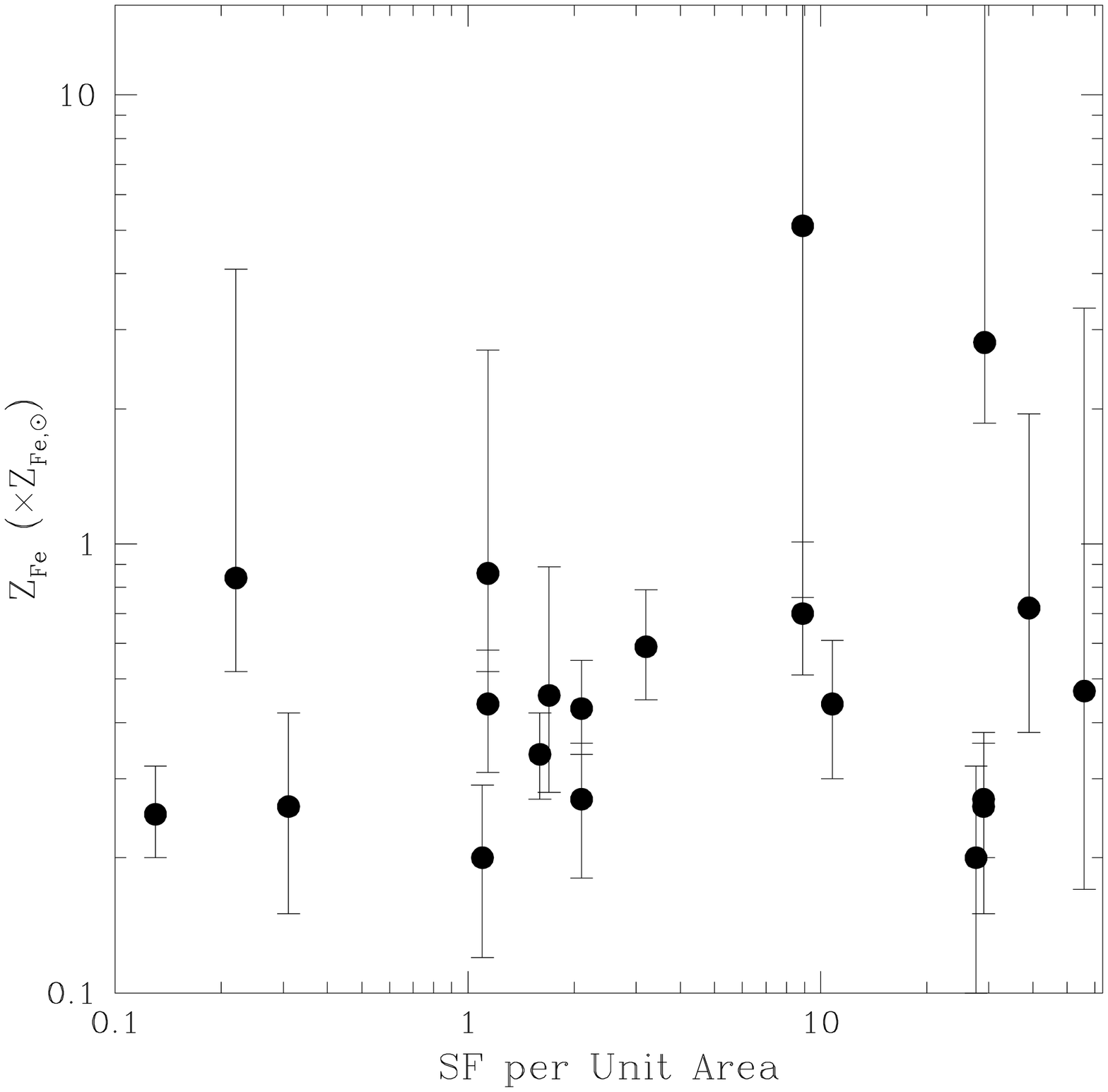}{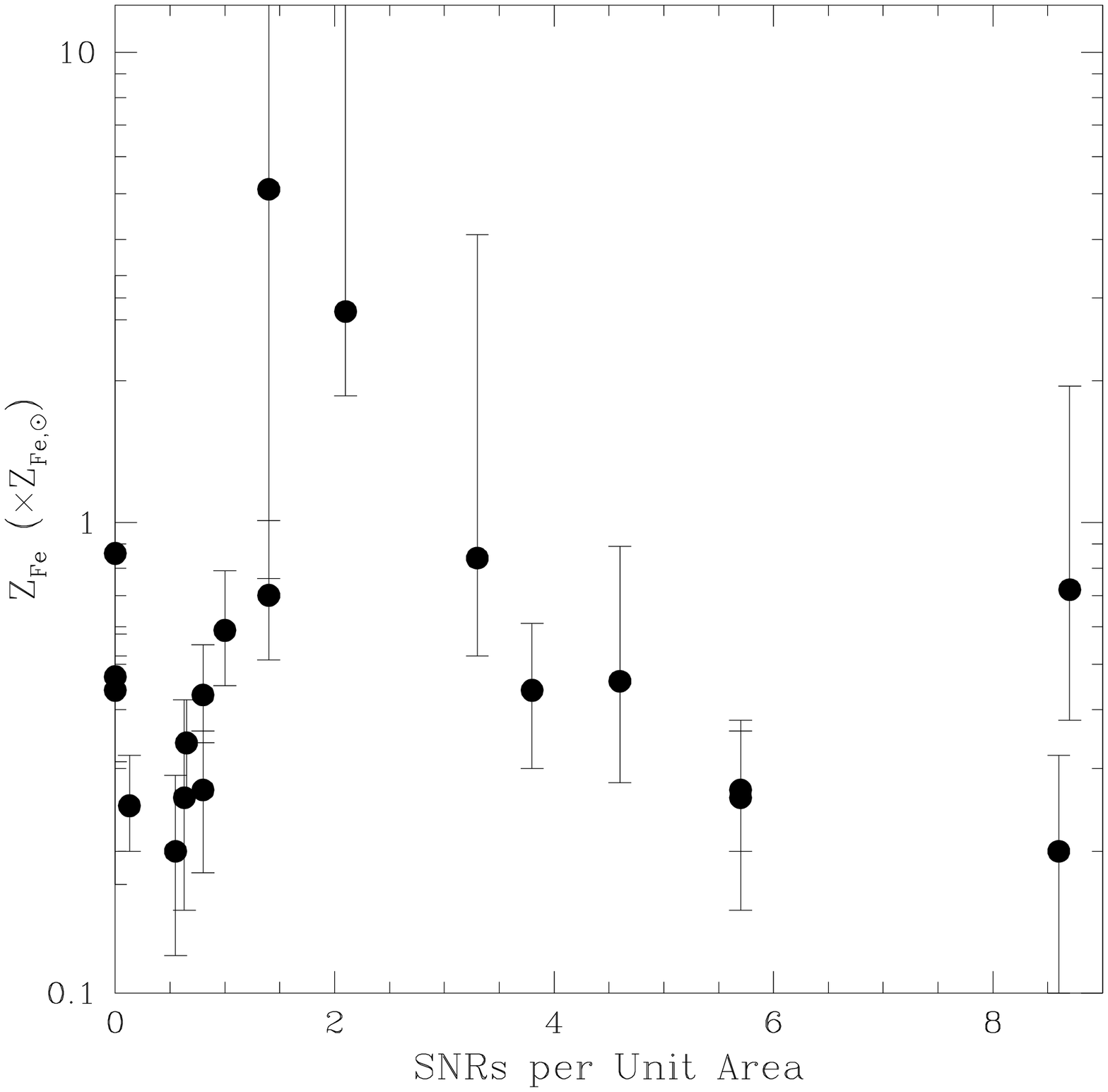}
\caption{Diagrams of the Fe abundance in each region of the Antennae 
hot ISM plotted versus (a) mean age of the star clusters, (b) relative strength
of star 
formation, (c) star formation per unit area, and (d) SNRs per unit area.
\label{optvsFe}}
\end{figure}

Figure~\ref{optvsNeMgSi} displays the relations between the four star-formation 
indicators listed above and the
mean abundance of $\alpha$-elements in each region. The mean abundance was 
calculated as the straight average
of the Ne, Mg, and Si abundances, while the error bars were computed 
through the propagation of errors
(at $1\sigma$).  As Figure~\ref{optvsNeMgSi} shows,
there are no strong correlations between the different star-formation 
indicators we calculated and the mean abundances in the same regions. 
The ``star-formation per unit area'' and the number of ``SNRs per unit area'' are supposed to be
among the best indicators of 
star formation.
Yet, these two quantities appear uncorrelated with the metal 
abundances, while a slight correlation may exist between the relative strength of star-formation 
and the abundance of $\alpha$-elements.
Figure~\ref{optvsFe} compares the same star-formation indicators with Fe abundance. 
No obvious trend
between the several quantities is detected in this case either, although 
a larger spread in Fe abundances seems to characterize regions with 
higher mean stellar ages. 

The hot-gas masses derived from abundances in \S~\ref{hotparam} may yield
some clues toward 
an interpretation of this lack of correlations. Assuming the average SN~II 
stellar yields listed in Nagataki \& Sato (1998), we estimated the number of 
type-II SNe necessary to produce the observed metal masses in the hot ISM, and 
the time required to produce that number of SNe.  If the resulting time is 
significantly shorter than $100$~Myr, then the lack of correlations may either be 
related to the speed at which mixing occurs in the hot ISM or indicate 
that the metals are somehow removed from the hot phase of the ISM. 
Conversely, if the SN production time is $\ga$100~Myr, then mean cluster 
ages may not be relevant because over periods of 100 Myr or longer major fractions of 
the decaying galaxy orbits are traversed and it is unclear whether---even if we were
able to derive better local mean ages (i.e., including also intermediate-age 
clusters)---any relation could be established between regional abundances and the past 
local star-formation history.

Table~\ref{snmetal} lists the results of our calculations for a range of 
theoretical models for SN~II average yields (reference given in column 1).
Columns (2) -- (5) give the numbers of type-II SNe necessary to produce the 
observed masses of 
Ne, Mg, Si, and Fe, respectively. From these numbers, we estimated the time
necessary to produce 
the required number of SNe II, assuming a constant SN rate throughout The Antennae. 
Neff \& Ulvestad (2000) derived a SN rate directly associated with the compact
radio sources detected in The Antennae of 0.03 yr$^{-1}$, assuming a typical SNR 
emits 6-cm radio emission for 30,000 yr and has the luminosity of Cas A.
However, applying the Condon \& Yin (1990) method (based on the ratio of non-thermal
radio luminosity to the radio-emitting SN rate in our Galaxy) to the total steep-spectrum
radio emission of The Antennae, and not only to the detected compact sources, they
predict a SN rate of 0.26 yr$^{-1}$.
The resulting times required to produce 
the quantity of Ne, Mg, Si, and Fe observed in The Antennae through SN II enrichment were
calculated
for both of these estimates of the SN rate and are listed 
in columns (6), (7), (8), and (9), respectively. For both rate estimates the times required 
are short: in the $\sim$3.4--12 Myr range for a SN rate of 0.03 yr$^{-1}$ and in the
$\sim$0.4--1.4 Myr range for 0.26 yr$^{-1}$.  Note that the second estimate
is more reliable because the SN rate was derived taking into account not only the 
radio emission from the identifiable compact sources, but also the overall 
steep-spectrum radio emission from the whole galaxy.

However, if the average density of the ISM is
lower than our estimates (e.g. if the disk thickness is larger than our assumed 
200~pc), so that more 
mass and more metals are needed to produce the observed 
line intensities, then the above times may increase to $\sim$3--4 Myr (or $\sim$25--35 Myr
in the lower SN rate estimate). Such time spans may be
too short for the metals to become mixed throughout the hot ISM.
Note also that a filling factor of the hot gas lower than 1 would increase the density
estimate and make such time spans shorter.

Indeed, 2 Myr is 
a relevant time threshold for mixing in a hot gas of the temperature we detect, 
corresponding to average velocities of $\sim 500$ km~s$^{-1}$.
At 500 km~s$^{-1}$, gas travels only $\sim1$ kpc in 2 Myr.
If the time 
required to produce the metals is actually of this order of $\la$2 Myr, 
it may be unlikely that efficient mixing takes place on a scale larger than 
a few hundred pc, and it is probably insufficient to destroy the correlation. 
Other factors may have destroyed the correlation as well
like, e.g., (i) the presence of a significant fraction of the SN ejecta in 
a cool phase or (ii) in dust grains, or (iii) the escape of the ejecta in a galactic wind.
Another possibility could be that the other metals are invisible in soft X-rays.
Indeed, in a Chevalier \& Clegg (1985) wind model the metals would be expected 
to be in much hotter ($T\sim10^{7.5}$ K) gas. Therefore the hot gas observed could be 
primarily ambient ISM heated, but not particularly enriched, by Type II SNe.

\section{Physical properties of the hot ISM}

The overall characteristics of the hot, diffuse gas in The
Antennae, as noted by Read et al.\ (1995) and F03, are (1) a very high pressure, 
ranging from
10 to 100 times the pressure of the hot component of
the ISM in the solar neighborhood
(Cox \& Reynolds 1987), (2) temperatures in the range 0.3 to 0.7 keV,
or about 3 to 7 times the temperature of the Galactic
hot ISM, and (3) an abundance of $\alpha$-process elements generally
consistent with the $\sim 0.5$ solar abundance shown by the stellar 
populations (Fritze-v.Alvensleben 1998). 

There are, of course, uncertainties both in the model
fits and in the path length that must
be assumed to convert emission measures to pressures.
In \S~\ref{hotparam} we assumed a 200 pc thickness for the
emitting region and a filling factor of 1, unless
explicit factors are given. Where two temperatures are
required, we assumed the components to be in pressure 
equilibrium and their combined path length
to be 200 pc. 

In the spectral 
analysis above we have pinned down the metallicity parameters for specific 
regions, and we find abundances of Ne, Mg, and Si
generally ranging from $\sim$0.2 $Z_\odot$ to $Z_\odot$,
along with a few cases of overabundance, represented mainly by
five regions: 4b, 5, 7, 8a and 8b.

Three of these regions can be defined as ``Hot Diffuse Emission
Regions''  (4b, 7, and 8a), requiring fitting either with
two temperatures, one of which consistently to be above 0.7 keV,
or with a single temperature above 0.5 keV.
On the
other hand, the ``Cooler Diffuse Emission Regions'' (5 and 8b)
are both well fit with a single temperature around 0.3 keV.

\subsection{Thermal Energy Supply from type II SNe}

The five regions overabundant in metals listed above represent an
heterogeneous subsample of our X-ray diffuse emission regions.

Region 4b lies in the X-ray ridge that runs from NW to SE.
Together with regions 4a, 6a and 6b, it is bounded to the NW by a strong H~I arm
and to the SE by a region of strong CO emission.
Star formation in this region is modest as judged
by the number of young clusters (Whitmore et al.\ 1999),
and radio continuum emission is correspondingly weak (Hibbard et al.\ 2001).
The mean age of the clusters is about 10 Myr (Whitmore \& Schweizer 1995).
Thus this region is relatively quiescent and presents
a relatively simple case for analysis.

On the other hand, Region 7 corresponds to the most active burst of star
formation currently occuring in the Antennae, as observed both in the optical
and in the near infrared (Mirabel et al.\ 1998; 
Wilson et al.\ 2000; Zhang et al.\ 2001). It also corresponds to
the ``Overlap Region'' between the two merging disks, and is discussed in
more detail by Fabbiano et al.\ (2004).

Regions 8a and 8b are strongly influenced by the presence of
the southern galactic nucleus and present, together with Region 7, the highest 
intrinsic absorptions measured in the various spectral fits.

Finally, Region 5 stands out for its very high abundances and high concentration
of young star clusters.  It, too, is discussed by Fabbiano et al.\ (2004).
Its high abundance of $\alpha$-elements compared to iron is consistent with 
enrichment by Type II SNe.  Its low temperature (see Table~\ref{mytable1}) could be the
result of a relatively high local gas density or of relatively effective
entrainment of cool ISM gas into the low density SN-heated gas.
It may evolve into a state similar to that of the neighboring regions
4 and 6 over the course of a few times $10^7$ years.

Following a similar procedure to the one described in \S~\ref{sectabund},
we derived first the mass of each chemical element from the spectral fits and
then the number of type II SNe required to produce 
the observed quantity of Ne, Mg, Si, and Fe in the five above regions.
The required number of SNe can yield an estimate of the total supply of thermal 
energy available to each region. Our estimates were performed by averaging
the numbers computed from each of the SN-yield models considered in Table~\ref{snmetal}
and propagating the errors.
Assuming an average energy supply of $10^{51}$ erg per SN, we then obtained the 
thermal energies listed in Table~\ref{eth}. 
For each region the thermal energy supply derived from the observed quantity of
Ne, Mg, Si, and Fe are shown in columns (2), (3), (4), and (5), respectively.

If we compare these numbers with the thermal energies derived from the spectral
fits (listed in Table~\ref{mytable2}), we can see that the observed thermal
energy is systematically lower than the energy supplied by SNe. 
Note that filling factors less than unity would not affect this
difference since both thermal energies and masses scale as
$\eta^{1/2}$. This 
discrepancy is generally about a factor of 5, with the exceptions of Regions
5 and 8b, where the observed thermal energy is lower by about 
two orders of magnitude.
Although the large errors in the metallicities of Region 8b can make this
discrepancy less dramatic (and comparable with the other three regions), Region 5
does not present such large uncertainties and clearly stands out with respect to 
the other emission regions. It is likely that in Region 5
most of the thermal energy of the SNe that enriched the hot gas ($\sim99\%$) 
has been lost either through thermal conduction and radiative cooling or 
through a wind.

Radiative cooling of the hot gas can be clearly ruled out by the cooling times
listed in Table~\ref{mytable2}, which are definitely too long.
On the other hand, the conductive cooling time scale for a volume of size $L$, 
temperature $T$, and density $n$ is
$$
t_{\rm cond} = \frac{5 nkT L^3}{10^{-5} T^{5/2} (T/L) L^2}
=   9\times10^8\;  n T_{0.5}^{-5/2} L_{\rm pc}^2\; {\rm sec}
$$
where $T_{0.5}$ is the temperature in units of 0.5 keV and $L_{\rm pc}$ is 
the length
scale in parsecs.  For $n\sim0.05$ cm$^{-3}$ (the typical electron density
we observe in The Antennae X-ray diffuse emission), a temperature of 0.5 keV,
and a length scale $L\sim30$ pc, this formula yields a conductive time scale 
of $\sim1000$ yr, assuming full Spitzer conductivity (Spitzer 1962). 
Thus, conduction may have been able to dissipate a significant fraction of 
the thermal energy produced by Type II SNe in Region 5 and in the other 
regions we examined, if hot and cool gas are mixed on a scale 
comparable to that in the Galaxy.

Loss of energy by a wind seems to be a possibility, too. However, a wind is supposed
to carry away metals as well. Therefore, one would require the energy loss to the wind to be
$\sim\,$100 times more efficient than the mass loss (at least for
Region 5). This would suggest a wind temperature $\sim\,$100 times higher than the 
temperature of the gas we observe, which looks implausible.

\section{Summary and Conclusions}

In this paper and in Paper I we have performed an extensive 
study of the X-ray properties of the diffuse 
emission of The Antennae (NGC~4038/39), analyzing the entire 411~ks exposure 
obtained with {\it Chandra} ACIS-S. 
Confirming the results of F03, based on the first of the seven 
observations used in the present study, we report a spatially and spectrally 
complex hot ISM. Thanks to our deep data, we also detect clear, spatially 
variable emission lines of Ne, Mg, and Si, in addition to the Fe-L blend, which
indicate super-solar metal abundances in a few regions of the hot ISM.
In this paper we have tried to investigate the origin of these metals.

To summarize:

1) We derived physical parameters of the hot gas, such as density, thermal energy,
cooling times and pressure.
This hot gas has cooling times lying in the $10^7$--$10^8$ yr range, 
in agreement with the results of F03.
At least in one of the two nuclei the hot-gas 
pressure is significantly higher than the CO pressure, implying that shock waves 
may be driven into the CO clouds.
Summing the hot-gas masses determined for each region yields a 
total measured mass of $\sim${}$1.2\times 10^7$ M$_\sun$ for the hot ISM.

3) Fitting the Fe-L, Ne-IX, Mg-XI and Si-XIII emission, we find significant 
metal enrichment of the hot ISM. Metal abundances are generally consistent with 
solar, but reach extremes of  $\sim 20-30$ solar in Region 5 and are significantly
subsolar in a few regions. 
The total hot gas masses derived for Fe, Ne, Mg, and Si are 
$1.8\times10^4$ M$_\sun$, $3.4\times10^4$ M$_\sun$, $1.3\times10^4$ M$_\sun$,
and $2.6\times10^4$ M$_\sun$, respectively.

4) Comparison of elemental ratios with those expected from Type~Ia and Type~II SNe suggests that 
SNe~II are mainly responsible for the metal enrichment of the hot ISM in The Antennae. However, 
we measure [Si/Fe] ratios mostly lower than those predicted for SN~II 
yields. A similar inconsistency was also found in Paper I, linking together all the $\alpha$ elements 
using the type II SNe ratio from the compilation of average stellar yields.
Although there surely are considerable uncertainties in the SN yields predicted by
theoretical models, it is also true that our spectra are affected by low statistics at the
energy of 
the Si line (E$\sim$2 keV). This suggests that uncertainties in the measurements
of the Si abundance may be the primary cause of this discrepancy. 
If our measurements are correct, our result may point to a depletion of Si in the 
hot ISM, perhaps from more efficient cooling of this element.

5) We report a remarkable lack of correlations between our abundance 
measurements and stellar age indicators, estimated from either the $Hubble$ 
observations of stellar clusters (Whitmore \& Schweizer 1995; Whitmore et al.\ 1999) or 
VLA SNR estimates (Neff \& Ulvestad 2000).
The time required to produce the observed quantities of metals through
type-II SN explosions is probably $\la$2 Myr, implying that 
efficient mixing is unlikely to be the main agent destroying the expected
correlation between abundances and radio-optical star-formation indicators. This
may point toward the presence of a significant fraction of SN~II ejecta 
in a cool phase, in dust grains, or escaping in a wind of hot gas, undetectable
at soft X-rays wavelengths.

6) We have analyzed the energy input coming from SN explosions in the 5 regions with
metallicities significantly higher than the average stellar abundance 
($\sim$0.5 solar; Fritze-v.Alvensleben 1998). In all of these regions 
we observe a thermal energy lower than expected from their metallicities
and SN enrichment.  We conclude that the missing
energy has been dissipated through processes different than radiative cooling
(radiative $\tau_c\sim10^7-10^8$ yr), most likely through conductive cooling.

%

\acknowledgements

We thank the {\it Chandra} X-ray Center DS and SDS teams for their efforts in
reducing the data and for developing the software used in the data reduction (SDP)
and analysis (CIAO).
We thank D.-W. Kim for useful discussions.
We also thank the anonymous referee for his careful analysis of the paper
and for giving us helpful suggestions for the presentation of our results.
This work was supported in part by NASA contract NAS8-39073 and NASA grants 
GO1-2115X and GO2-3135X.
F.S. acknowledges partial support from the NSF through grant AST 02-05994.

\clearpage

\begin{figure}[H]
\epsscale{0.85}
\caption{Mapped-color broad-band smoothed image of The Antennae. The red
color channel indicates emission from the 0.3--0.65 keV band, green from the
0.65--1.5 keV band, and blue from the 1.5--6 keV band.
\label{truecolor}}
\end{figure}


\begin{figure}[H]
\epsscale{0.90}
\caption{Mapped-color line-strength image of The Antennae. Red represents
emission from the Fe K-line blend (plus O and Ne), green from the Mg line, and blue 
from the Si line (for details, see text). The 17 regions used for a preliminary 
subdivision of the hot ISM (performed in Paper I) are marked in white.
\label{linemap}}
\end{figure}


\begin{figure}[H]
\epsscale{0.90}
\caption{0.3--6 keV image of The Antennae, obtained after removing the $\ge$3$\sigma$ 
point sources.
The 21 regions used for the spectral analysis of the hot ISM are marked
in white.
\label{regions}}
\end{figure}

\clearpage

\clearpage 

\clearpage 

\clearpage 

\begin{deluxetable}{cccccccccccc}
\tabletypesize{\scriptsize}
\rotate
\tablecaption{Best-Fit Parameters (1$\sigma$ errors for all interesting parameters, see text for details) 
for the Diffuse Emission of The Antennae. 
\label{mytable1}}
\tablewidth{0pt}
\tablehead{
\colhead{Region \#} &
\colhead{$\chi^2/dof$} &
\colhead{\begin{tabular}{c}
$N_H$\\
($\times10^{20}$ cm$^{-2}$)
\end{tabular}} &
\colhead{\begin{tabular}{c}
$kT$\\
(keV)
\end{tabular}} &
\colhead{\begin{tabular}{c}
$Z_{Ne}$\\
($\times Z_{Ne,\odot}$)
\end{tabular}} &
\colhead{\begin{tabular}{c}
$Z_{Mg}$\\
($\times Z_{Mg,\odot}$)
\end{tabular}} &
\colhead{\begin{tabular}{c}
$Z_{Si}$\\
($\times Z_{Si,\odot}$)
\end{tabular}} &
\colhead{\begin{tabular}{c}
$Z_{Fe}$\\
($\times Z_{Fe,\odot}$)
\end{tabular}} &
\colhead{$\Gamma$} &
\colhead{Mod\tablenotemark{(*)}}
}
\startdata
1&43.5/38&$<8.60$&$0.61_{-0.10}^{+0.07}$&$<0.75$&$0.15_{-0.15}^{+0.47}$&$<0.51$&$0.20_{-0.08}^{+0.09}$&\nodata&A1\\
\hline
2&34.3/25&$<4.95$&$0.62_{-0.18}^{+0.10}$&$0.20_{-0.20}^{+0.84}$&$0.35_{-0.35}^{+0.61}$&$0.19_{-0.19}^{+2.69}$&$0.20_{-0.10}^{+0.12}$&\nodata&A1\\
\hline
3&57.8/48&$<6.73$&$0.54_{-0.14}^{+0.09}$&$0.48_{-0.48}^{+1.01}$&$0.56_{-0.54}^{+0.85}$&$0.51_{-0.51}^{+1.80}$&$0.26_{-0.11}^{+0.16}$&\nodata&A1\\
\hline
4a&51.1/44&$1.42_{-1.42}^{+8.10}$&$0.63_{-0.06}^{+0.07}$&$0.38_{-0.38}^{+0.78}$&$0.86_{-0.65}^{+0.76}$&$0.23_{-0.23}^{+0.56}$&$0.44_{-0.13}^{+0.14}$&
\nodata&A1\\
\hline
4b&88.2/63&$0.97_{-0.97}^{+4.22}$&\begin{tabular}{c}
$0.20 (unconstr.)$\\ 
$>0.54$
\end{tabular}
&$1.35_{-0.90}^{+1.46}$&$1.44_{-0.84}^{+3.93}$&$1.21_{-0.80}^{+1.13}$&$0.86_{-0.34}^{+1.84}$&\nodata&A2\\
\hline
5&31.4/35&$0.91_{-0.91}^{+10.69}$&$0.30_{-0.08}^{+0.07}$&$8.86_{-4.24}^{+12.47}$&
$19.07_{-14.08}^{+293.43}$&$109.82_{-109.82}^{+510.25}$&$2.81_{-0.95}^{+87.43}$&
$1.88$\tablenotemark{(+)}
&B1\tablenotemark{(+)}\\
\hline
6a&74.6/61&$0.74_{-0.74}^{+5.68}$&$0.66_{-0.07}^{+0.06}$&$0.16_{-0.16}^{+0.56}$&$0.28_{-0.28}^{+0.43}$&$0.26_{-0.26}^{+0.47}$&$0.27\pm0.09$&
$-0.61_{-0.39}^{+1.15}$&B1\\
\hline
6b&95.5/80&$1.47_{-1.47}^{+1.79}$&$0.61\pm0.04$&$0.72_{-0.39}^{+0.44}$&$0.94_{-0.37}^{+0.49}$&$0.80_{-0.38}^{+0.46}$&$0.43_{-0.09}^{+0.12}$&\nodata&A1\\
\hline
7&86.9/65&$13.77_{-9.39}^{+11.69}$&$0.56_{-0.16}^{+0.12}$&$2.35_{-1.83}^{+3.85}$&$1.93_{-1.19}^{+3.00}$&$2.74_{-1.77}^{+3.02}$&
$0.46_{-0.18}^{+0.43}$&$1.88$\tablenotemark{(+)}&B1\tablenotemark{(+)}\\
\hline
8a&102.5/90&$8.66_{-4.81}^{+5.94}$&$0.61_{-0.05}^{+0.04}$&$1.50_{-0.87}^{+1.65}$&$0.82_{-0.53}^{+0.88}$&$0.61_{-0.44}^{+0.78}$&
$0.70_{-0.19}^{+0.31}$&$1.88$\tablenotemark{(+)}&B1\tablenotemark{(+)}\\
\hline
8b&82.5/71&$24.61_{-11.43}^{+11.60}$&$0.33_{-0.05}^{+0.28}$&$15.57_{-14.62}^{+\infty}$&$5.95_{-5.33}^{+132.59}$&$12.44_{-12.04}^{+\infty}$&
$5.11_{-4.35}^{+63.59}$&$3.61_{-0.70}^{+0.64}$&B1\\
\hline
9&120.6/108&$2.82_{-2.82}^{+3.86}$&$0.62_{-0.05}^{+0.04}$&$1.26_{-0.67}^{+0.99}$&$0.76_{-0.43}^{+0.55}$&$0.84_{-0.46}^{+0.60}$&
$0.59_{-0.14}^{+0.20}$&$1.88$\tablenotemark{(+)}&B1\tablenotemark{(+)}\\
\hline
10&128.2/93&$<2.04$&$0.63_{-0.05}^{+0.04}$&$0.33_{-0.33}^{+0.41}$&$0.41_{-0.28}^{+0.31}$&$0.70_{-0.41}^{+0.56}$&$0.25_{-0.05}^{+0.07}$&\nodata&A1\\
\hline
11&57.2/54&$1.88_{-1.88}^{+7.82}$&$0.62\pm0.06$&$0.59_{-0.59}^{+0.95}$&$1.27_{-0.75}^{+0.94}$&$0.87_{-0.63}^{+0.93}$&$0.44_{-0.14}^{+0.17}$&\nodata&A1\\
\hline
12a&55.0/46&$<2.30$&$0.60_{-0.09}^{+0.05}$&$0.11_{-0.11}^{+0.48}$&$0.35_{-0.32}^{+0.39}$&$0.20_{-0.20}^{+0.65}$&$0.27_{-0.07}^{+0.09}$&\nodata&A1\\
\hline
12b&24.5/28&$<5.29$&$0.61\pm0.06$&$<0.38$&$0.14_{-0.14}^{+0.65}$&$<0.82$&$0.26_{-0.11}^{+0.12}$&\nodata&A1\\
\hline
13&32.2/22&$0.32_{-0.32}^{+13.04}$&$0.58_{-0.35}^{+0.09}$&$2.57_{-2.57}^{+5.42}$&$1.33_{-1.26}^{+15.18}$&$1.54_{-1.54}^{+136.71}$&$0.47_{-0.30}^{+2.88}$&
$1.88$\tablenotemark{(+)}&B1\tablenotemark{(+)}\\
\hline
14&28.8/21&$3.07_{-3.07}^{+29.18}$&$0.37_{-0.08}^{+0.39}$&$unconstr.$&$unconstr.$&$unconstr.$&$unconstr.$&$1.88$\tablenotemark{(+)}&
B1\tablenotemark{(+)}\\
\hline
15&44.9/38&$3.95_{-3.95}^{+10.92}$&$0.59_{-0.14}^{+0.09}$&$1.09_{-1.09}^{+5.52}$&$0.89_{-0.89}^{+2.37}$&$0.37_{-0.37}^{+1.61}$&
$0.72_{-0.34}^{+1.23}$&$1.88$\tablenotemark{(+)}&B1\tablenotemark{(+)}\\
\hline
16&67.0/78&$0.76_{-0.76}^{+4.07}$&$0.62_{-0.05}^{+0.04}$&$0.32_{-0.32}^{+0.46}$&$0.52_{-0.32}^{+0.38}$&$0.56_{-0.34}^{+0.44}$&$0.34_{-0.07}^{+0.08}$&
\nodata&A1\\
\hline
17&90.5/73&$3.08_{-3.08}^{+10.26}$&$0.62_{-0.34}^{+0.10}$&$0.96_{-0.96}^{+14.74}$&$2.91_{-2.24}^{+28.48}$&$1.16_{-1.16}^{+4.94}$&$0.84_{-0.32}^{+3.25}$&
$1.88$\tablenotemark{(+)}&B1\tablenotemark{(+)}\\
\enddata

\tablenotetext{(*)}{XSPEC Models: A1 = {\em wabs(wabs(vapec))}; A2 = {\em 
wabs(wabs(vapec+vapec))}; 
B1 = {\em wabs(wabs(vapec+powerlaw))}; B2 = 
{\em wabs(wabs(vapec+vapec+powerlaw))}.}
\tablenotetext{(+)}{Photon index frozen.}

\end{deluxetable}
\clearpage


\begin{deluxetable}{ccccccc}
\rotate
\tabletypesize{\footnotesize}
\tablecaption{Emission Parameters for the Diffuse Emission of The 
Antennae.\label{lumin}}
\tablewidth{0pt}
\tablehead{
\colhead{\begin{tabular}{c}
Region \#\\
(1)
\end{tabular}} &
\colhead{\begin{tabular}{c}
$F_{0.3-10\:keV}^{p.l.}$\\
($\times10^{-14}$ cgs)\\
(2)
\end{tabular}} &
\colhead{\begin{tabular}{c}
$F_{0.3-10\:keV}^{therm}$\\
($\times10^{-14}$ cgs)\\
(3)
\end{tabular}} &
\colhead{\begin{tabular}{c}
$EM$\tablenotemark{(*)}\\
($\times10^{61}$ cm$^{-3}$)\\
(4)
\end{tabular}} &
\colhead{\begin{tabular}{c}
$L_{0.3-10\:keV}^{p.l.}$\\
($\times10^{38}$ cgs)\\
(5)
\end{tabular}} &
\colhead{\begin{tabular}{c}
$L_{0.3-10\:keV}^{therm}$\\
($\times10^{38}$ cgs)\\
(6)
\end{tabular}} &
\colhead{\begin{tabular}{c}
$\frac{L_{0.3-10\:keV}^{p.l.}}{L_{0.3-10\:keV}^{therm}}$\\
(7)
\end{tabular}} 
}
\startdata
1 & \nodata & $0.80_{-0.16}^{+0.53}$ & $3.79_{-0.74}^{+2.49}$ & \nodata & $3.47_{-0.67}^{+2.30}$ & \nodata \\
2 & \nodata & $0.65_{-0.23}^{+0.26}$ & $2.85_{-1.03}^{+1.11}$ & \nodata & $2.81_{-0.97}^{+1.10}$ & \nodata \\
3 & \nodata & $0.87_{-0.32}^{+0.49}$ & $3.65_{-1.31}^{+2.02}$ & \nodata & $3.78_{-1.37}^{+2.05}$ & \nodata \\
4a & \nodata & $1.14_{-0.29}^{+0.89}$ & $4.08_{-1.05}^{+3.20}$ & \nodata & $5.23_{-1.33}^{+4.09}$ & \nodata \\
4b\tablenotemark{(+)} & \nodata & $0.27_{-0.22}^{+0.32}$, $2.27_{-1.55}^{+1.65}$ & $0.97$, $4.17$ & \nodata & $1.27_{-1.01}^{+1.51}$, $10.2_{-7.0}^{+7.4}$ & \nodata \\
5 & $0.44\pm0.21$ & $0.76_{-0.43}^{+0.92}$ & $0.85_{-0.48}^{+1.03}$ & $1.93_{-0.93}^{+0.94}$ & $3.41_{-1.94}^{+4.13}$ & 0.566 \\
6a & $1.68_{-1.31}^{+32.44}$ & $1.92_{-0.30}^{+0.90}$ & $8.41_{-1.43}^{+3.87}$ & $7.26_{-5.66}^{+140.12}$ & $8.57_{-1.33}^{+4.02}$ & 0.847 \\
6b & \nodata & $3.48_{-0.71}^{+1.18}$ & $11.4_{-2.4}^{+3.8}$ & \nodata & $16.0_{-3.2}^{+5.4}$ & \nodata \\
7 & $1.21_{-0.50}^{+0.42}$ & $0.91_{-0.53}^{+2.38}$ & $2.76_{-1.61}^{+7.20}$ & $6.37_{-2.63}^{+2.21}$ & $6.10_{-3.54}^{+15.98}$ & 1.044 \\
8a & $1.63_{-0.49}^{+0.46}$ & $3.15_{-0.73}^{+1.72}$ & $10.6_{-2.5}^{+5.7}$ & $8.12_{-2.44}^{+2.29}$ & $18.7_{-4.3}^{+10.2}$ & 0.434 \\
8b & $0.87_{-0.40}^{+0.64}$ & $2.38_{-1.37}^{+7.99}$ & $2.08_{-1.19}^{+6.96}$ & $6.10_{-2.81}^{+4.51}$ & $14.8_{-8.5}^{+49.4}$ & 0.412 \\
9 & $1.94_{-0.64}^{+0.61}$ & $4.46_{-1.13}^{+2.25}$ & $12.4_{-3.1}^{+6.3}$ & $8.87_{-2.92}^{+2.79}$ & $21.6_{-5.5}^{+10.8}$ & 0.411 \\
10 & \nodata & $2.86_{-0.43}^{+0.55}$ & $11.2_{-1.7}^{+2.1}$ & \nodata & $12.4_{-1.9}^{+2.3}$ & \nodata \\
11 & \nodata & $1.46_{-0.41}^{+1.21}$ & $4.76_{-1.35}^{+3.92}$ & \nodata & $6.81_{-1.91}^{+5.61}$ & \nodata \\
12a & \nodata & $1.49_{-0.27}^{+0.33}$ & $6.24_{-1.06}^{+1.36}$ & \nodata & $6.44_{-1.17}^{+1.42}$ & \nodata \\
12b & \nodata & $0.74_{-0.17}^{+0.32}$ & $3.11_{-0.70}^{+1.36}$ & \nodata & $3.19_{-0.72}^{+1.37}$ & \nodata \\
13 & $0.16_{-0.16}^{+0.35}$ & $0.55_{-0.41}^{+0.66}$ & $1.91_{-1.43}^{+2.35}$ & $0.68_{-0.68}^{+1.51}$ & $2.40_{-1.80}^{+2.87}$ & 0.283 \\
14 & $0.25_{-0.17}^{+0.19}$ & $0.49_{-0.29}^{+3.68}$ & $0.09_{-0.05}^{+0.67}$ & $1.09_{-0.74}^{+0.83}$ & $2.12_{-1.26}^{+15.90}$ & 0.514 \\
15 & $0.59_{-0.33}^{+0.29}$ & $0.81_{-0.67}^{+1.08}$ & $2.17_{-1.79}^{+2.92}$ & $2.75_{-1.55}^{+1.36}$ & $4.08_{-3.34}^{+5.45}$ & 0.674 \\
16 & \nodata & $2.81_{-0.46}^{+1.02}$ & $10.3_{-1.7}^{+3.8}$ & \nodata & $12.5_{-2.0}^{+4.5}$ & \nodata \\
17 & $1.01_{-0.50}^{+0.47}$ & $0.96_{-0.58}^{+2.06}$ & $2.03_{-1.22}^{+4.32}$ & $4.64_{-2.30}^{+2.16}$ & $4.65_{-2.80}^{+9.93}$ & 0.998 \\
\enddata
\tablenotetext{(*)}{Emission measure defined as $EM=n^2V$.}
\tablenotetext{(+)}{In the case of two-temperature gas, the quantities are listed separately for both components: 
lower and higher temperature, respectively. For the calculation of $EM$ we assumed the best-fit value of $kT_1$ and 
$kT_2$; it was not possible to determine a confidence range for $EM$ because of the indetermination in both temperatures}

\end{deluxetable}

\clearpage


\begin{deluxetable}{cccccc}
\rotate
\tabletypesize{\footnotesize}
\tablecaption{Hot-Gas Parameters.\label{mytable2}}
\tablewidth{0pt}
\tablehead{
\colhead{\begin{tabular}{c}
Reg \#\\
(1)
\end{tabular}} &
\colhead{\begin{tabular}{c}
Area\\
(kpc$^2$)\\
(2)
\end{tabular}} &
\colhead{\begin{tabular}{c}
$n_e$\\
($\times10^{-2}$ cm$^{-3}$)\\
(3)
\end{tabular}} &
\colhead{\begin{tabular}{c}
$E_{th}$\\
($\times10^{53}$ erg)\\
(4)
\end{tabular}} &
\colhead{\begin{tabular}{c}
$\tau_c$\\
($\times10^{7}$ yr)\\
(5)
\end{tabular}} &
\colhead{\begin{tabular}{c}
$p$\\
($\times10^{-11}$ dyne cm$^{-2}$)\\
(6)
\end{tabular}}
}
\startdata
1 & 2.260 & $5.34_{-0.55}^{+1.54}$ & $20.8_{-5.2}^{+9.0}$ & $19.0_{-10.4}^{+14.7}$ & $10.4_{-2.6}^{+4.6}$ \\
2 & 0.597 & $9.01_{-1.81}^{+1.61}$ & $9.41_{-4.08}^{+3.46}$ & $10.6_{-6.3}^{+11.6}$ & $17.9_{-7.8}^{+6.6}$ \\
3 & 3.361 & $4.30_{-0.86}^{+1.06}$ & $22.0_{-9.0}^{+10.0}$ & $18.4_{-11.3}^{+23.7}$ & $7.43_{-3.03}^{+3.38}$ \\
4a & 0.960 & $8.50_{-1.17}^{+2.86}$ & $14.5_{-3.2}^{+7.0}$ & $8.79_{-4.95}^{+8.68}$ & $17.1_{-3.7}^{+8.3}$ \\
4b\tablenotemark{(*)} & 1.453 & $21.2$, $7.08$ & $0.44$, $17.0$ & $1.10$, $5.28$ & $13.6$ \\
5 & 0.483 & $5.48_{-1.88}^{+2.67}$ & $2.24_{-1.16}^{+1.86}$ & $2.08_{-1.63}^{+6.76}$ & $5.26_{-2.73}^{+4.39}$ \\
6a & 1.405 & $10.1_{-0.9}^{+2.1}$ & $26.4_{-4.9}^{+8.4}$ & $9.76_{-4.35}^{+5.47}$ & $21.3_{-3.9}^{+6.8}$ \\
6b & 2.551 & $8.71_{-0.95}^{+1.36}$ & $38.2_{-6.4}^{+8.9}$ & $7.57_{-2.86}^{+4.09}$ & $17.0_{-2.9}^{+3.9}$ \\
7 & 2.351 & $4.47_{-1.58}^{+4.02}$ & $16.6_{-8.9}^{+21.7}$ & $8.62_{-7.51}^{+38.80}$ & $8.01_{-4.31}^{+10.46}$ \\
8a & 0.856 & $14.5_{-1.8}^{+3.5}$ & $21.4_{-4.2}^{+6.8}$ & $3.63_{-1.74}^{+2.58}$ & $28.3_{-5.5}^{+9.1}$ \\
8b & 0.756 & $6.84_{-2.37}^{+7.43}$ & $4.81_{-2.67}^{+13.75}$ & $1.03_{-0.92}^{+8.31}$ & $7.22_{-3.21}^{+20.64}$ \\
9 & 4.528 & $6.84_{-0.93}^{+1.55}$ & $54.2_{-11.2}^{+16.5}$ & $7.95_{-3.74}^{+5.97}$ & $13.6_{-2.8}^{+4.1}$ \\
10 & 7.783 & $4.94_{-0.39}^{+0.45}$ & $68.3_{-10.4}^{+11.0}$ & $17.5_{-5.0}^{+6.4}$ & $9.96_{-1.52}^{+1.60}$ \\
11 & 1.317 & $7.84_{-1.20}^{+2.75}$ & $18.0_{-4.2}^{+8.7}$ & $8.38_{-4.86}^{+8.89}$ & $15.6_{-3.7}^{+7.4}$ \\
12a & 0.583 & $13.5_{-1.2}^{+1.4}$ & $13.3_{-3.0}^{+2.6}$ & $6.54_{-2.39}^{+3.02}$ & $25.9_{-5.8}^{+5.1}$ \\
12b & 0.266 & $14.1_{-1.7}^{+2.8}$ & $6.46_{-1.34}^{+2.04}$ & $6.42_{-2.86}^{+4.49}$ & $27.5_{-5.7}^{+8.7}$ \\
13 & 0.269 & $11.0_{-5.5}^{+5.4}$ & $4.85_{-3.89}^{+3.50}$ & $6.40_{-5.82}^{+37.70}$ & $20.4_{-16.4}^{+14.8}$ \\
14 & 0.171 & $2.98_{-1.08}^{+5.70}$ & $1.43_{-1.16}^{+1.74}$ & $2.14_{-2.09}^{+9.54}$ & $3.53_{-1.77}^{+17.58}$ \\
15 & 0.261 & $11.9_{-6.9}^{+6.3}$ & $3.25_{-1.59}^{+5.87}$ & $2.52_{-1.97}^{+36.50}$ & $22.5_{-15.3}^{+17.1}$ \\
16 & 3.391 & $7.20_{-0.62}^{+1.21}$ & $42.7_{-6.8}^{+10.4}$ & $10.8_{-4.1}^{+5.2}$ & $14.3_{-2.3}^{+3.5}$ \\
17 & 5.363 & $2.54_{-0.94}^{+1.95}$ & $23.8_{-17.0}^{+25.1}$ & $16.2_{-14.7}^{+67.6}$ & $5.04_{-3.61}^{+5.30}$ \\
\enddata

\tablenotetext{(*)}{In the case of two-temperature gas, we calculated $n_e$, 
$E_{th}$, and $\tau_c$ for both components (assuming the best fit temperatures and
pressure equilibrium): 
lower and higher temperature, respectively. It was not possible to determine a confidence 
range for those parameters and for $p$ because of the indetermination in both temperatures.}

\end{deluxetable}

\clearpage


\begin{deluxetable}{cccccc}
\tablecaption{Hot-Gas Masses for Hydrogen and Heavy Elements.\label{masses}}
\tabletypesize{\small}
\tablewidth{0pt}
\tablehead{
\colhead{Reg \#} &
\colhead{\begin{tabular}{c}
Total Mass\\
($M_\odot$)
\end{tabular}} &
\colhead{\begin{tabular}{c}
Ne Mass\\
($M_\odot$)
\end{tabular}} &
\colhead{\begin{tabular}{c}
Mg Mass\\
($M_\odot$)
\end{tabular}} &
\colhead{\begin{tabular}{c}
Si Mass\\
($M_\odot$)
\end{tabular}} &
\colhead{\begin{tabular}{c}
Fe Mass\\
($M_\odot$)
\end{tabular}} 
}
\startdata
1  & $5.9\times10^5$ & $<1.1\times10^3$ & $8.2\times10^1$ & $<3.0\times10^2$ & 
$3.1\times10^2$\\
2  & $2.6\times10^5$ & $1.3\times10^2$ & $8.5\times10^1$ & $5.0\times10^1$ & 
$1.4\times10^2$\\
3  & $7.1\times10^5$ & $8.5\times10^2$ & $3.7\times10^2$ & $3.6\times10^2$ & 
$4.8\times10^2$\\
4a & $4.0\times10^5$ & $3.8\times10^2$ & $3.2\times10^2$ & $9.2\times10^1$ & 
$4.6\times10^2$\\
4b & $5.3\times10^5$ & $1.8\times10^3$ & $7.1\times10^2$ & $6.4\times10^2$ & 
$1.2\times10^3$\\
5  & $1.3\times10^5$ & $2.9\times10^3$ & $2.3\times10^3$ & $1.4\times10^4$ & 
$9.5\times10^2$\\
6a & $7.0\times10^5$ & $2.8\times10^2$ & $1.8\times10^2$ & $1.8\times10^2$ & 
$4.9\times10^2$\\
6b & $1.1\times10^6$ & $2.0\times10^3$ & $9.5\times10^2$ & $8.7\times10^2$ & 
$1.2\times10^3$\\
7  & $5.2\times10^5$ & $3.0\times10^3$ & $9.2\times10^2$ & $1.4\times10^3$ & 
$6.2\times10^2$\\
8a & $6.1\times10^5$ & $2.3\times10^3$ & $4.6\times10^2$ & $3.7\times10^2$ & 
$1.1\times10^3$\\
8b & $2.5\times10^5$ & $9.8\times10^3$ & $1.4\times10^3$ & $3.2\times10^3$ & 
$3.4\times10^3$\\
9  & $1.5\times10^6$ & $4.8\times10^3$ & $1.1\times10^3$ & $1.3\times10^3$ & 
$2.3\times10^3$\\
10 & $1.9\times10^6$ & $1.6\times10^3$ & $7.2\times10^2$ & $1.3\times10^3$ & 
$1.2\times10^3$\\
11 & $5.1\times10^5$ & $7.4\times10^2$ & $6.0\times10^2$ & $4.4\times10^2$ & 
$5.8\times10^2$\\
12a & $3.9\times10^5$ & $1.1\times10^2$ & $1.3\times10^2$ & $7.7\times10^1$ & 
$2.7\times10^2$\\
12b & $1.8\times10^5$ & $<1.7\times10^2$ & $2.4\times10^1$ & $<1.5\times10^2$ & 
$1.3\times10^2$\\
13 & $1.5\times10^5$ & $9.3\times10^2$ & $1.8\times10^2$ & $2.2\times10^2$ & 
$1.8\times10^2$\\
14 & $2.5\times10^4$ & \nodata & \nodata & \nodata & \nodata\\
15 & $1.5\times10^5$ & $4.1\times10^2$ & $1.3\times10^2$ & $5.6\times10^1$ & 
$2.9\times10^2$\\
16 & $1.2\times10^6$ & $9.5\times10^2$ & $5.8\times10^2$ & $6.7\times10^2$ & 
$1.1\times10^3$\\
17 & $6.7\times10^5$ & $1.6\times10^3$ & $1.8\times10^3$ & $7.7\times10^2$ & 
$1.5\times10^3$\\
\enddata

\end{deluxetable}

\clearpage


\begin{deluxetable}{cccccccccc}
\tablecaption{Mean Optical and Radio Observed Properties of the 17 {\it Chandra} Regions in 
NGC~4038/39.\label{francois}}
\tabletypesize{\footnotesize}
\rotate
\tablewidth{0pt}
\tablehead{
\colhead{Reg\#} &
\colhead{\begin{tabular}{c}
HST\\
WFPC2\\
$\#R$
\end{tabular}} &
\colhead{\begin{tabular}{c}
HST\\
WFPC2\\
$\#B$
\end{tabular}} &
\colhead{\begin{tabular}{c}
HST\\
WFPC2\\
$\#I$
\end{tabular}} &
\colhead{\begin{tabular}{c}
VLA\\
$fl$
\end{tabular}} &
\colhead{\begin{tabular}{c}
VLA\\
$st$
\end{tabular}} &
\colhead{\begin{tabular}{c}
Phot\\
HST\\
$U$
\end{tabular}} &
\colhead{\begin{tabular}{c}
Phot\\
HST\\
$V$
\end{tabular}} &
\colhead{\begin{tabular}{c}
Phot\\
HST\\
$I$
\end{tabular}} &
\colhead{\begin{tabular}{c}
Phot\\
HST\\
$H\alpha$
\end{tabular}}
}
\startdata
 1\tablenotemark{(a)} & 1 & 0 & 0 & 0 & 1 & 1 & 2 & 3 & 1 \\
 2 & 1 & 3 & 0 & .5 & 5 & 9 & 5 & 4 & 7 \\
 3 & 2 & 0 & 0 & 0 & 2 & 0 & 2 & 4 & 1 \\
 4 & 1 & 2 & 2 & 0 & 0 & 1 & 3 & 6 & 2 \\
 5 & 2 & 3 & 0 & 2 & 1 & 6 & 3 & 3 & 8 \\
 6 & 5 & 6 & 2 & 0 & 3 & 4 & 2 & 3 & 4 \\
 7 & 16 & 0 & 0 & 3 & 11 & 0 & 1 & 3 & 4 \\
 8 & 2 & 1 & 0 & 1 & 2 & 3 & 6 & 10 & 10 \\
 9 & 0 & 7 & 0 & 3 & 4 & 3 & 4 & 5 & 10 \\
10 & 0 & 0 & 1 & 0 & 1 & 0 & 0 & 0 & 1 \\
11 & 4 & 3 & 2 & 1 & 5 & 6 & 3 & 3 & 8 \\
12 & 2 & 6 & 1 & 1 & 4 & 10 & 6 & 5 & 10 \\
13 & 0 & 3 & 0 & 2 & 0 & 6 & 5 & 5 & 8 \\
14 & 1 & 0 & 0 & 0 & 1 & 1 & 6 & 10 & 5 \\
15 & 0 & 2 & 0 & 0 & 2 & 3 & 3 & 8 & 6 \\
16 & 2 & 2 & 0 & 1 & 2 & 2 & 3 & 5 & 3 \\
17 & 10 & 2 & 1 & 3 & 15 & 0 & 2 & 3 & 1 
\enddata

\tablenotetext{(a)}{Values for Region 1 based only on eastern part
(western covered by WFPC2 to $<50\%$)}

\end{deluxetable}

\clearpage



\begin{deluxetable}{cccccccl}
\tablecaption{Mean Optical and Radio Derived Properties of the 17 {\it Chandra} Regions in 
NGC~4038/39.\label{francois2}}
\tabletypesize{\scriptsize}
\rotate
\tablewidth{0pt}
\tablehead{
\colhead{Reg\#} &
\colhead{\begin{tabular}{c}
Mean\\
Age\\
(Myr)
\end{tabular}} &
\colhead{\begin{tabular}{c}
Relative\\
Strength\\
SF
\end{tabular}} &
\colhead{\begin{tabular}{c}
SF per\\
Unit\\
Area
\end{tabular}} &
\colhead{\begin{tabular}{c}
SNRs\\
per\\
Area
\end{tabular}} &
\colhead{\begin{tabular}{c}
Unit\\
Area\\
(kpc$^2$)
\end{tabular}} &
\colhead{Gal} &
\colhead{Landmark Objects}
}
\startdata
 1\tablenotemark{(a)} & 4.0 & 0.4 & 1.10 & 0.55 & 1.81 & 38 & North in ``W Loop''\\
 2 & 12.2 & 1.8 & 27.6 & 8.6 & 0.58 & 38 & ``G'' complex \\
 3 & 4.0 & 0.2 & 0.31 & 0.63 & 3.2 & 38 & Nearly pure old disk \\
 4 & 11.3 & 0.3 & 1.14 & 0.0 & 2.64 & 38 & ``H'' \& outer bulge, dust lane \\
 5 & 10.6 & 2.3 & 29.2 & 2.1 & 0.48 & 38 & 2/3 ``F,'' disk \\
 6 & 10.0 & 1.6 & 2.1 & 0.8 & 3.8 & 38 & 2/3 ``E,'' strong SF in NE disk \\
 7 & 4.0 & 1.0 & 1.7 & 4.6 & 2.4 & \nodata & ``Overlap region'' \\
 8 & 7.7 & 0.8 & 8.9 & 1.4 & 1.46 & 39 & ``A,'' nucleus \& bulge of 39 \\
 9 & 15.0 & 1.4 & 3.2 & 1.0 & 4.0 & 39 & ``B''+``AA,'' outer bulge of 39 \\
10 & \nodata & \nodata & 0.13 & 0.13 & 7.8 & \nodata & Interdisk region, $\sim$empty \\
11 & 8.7 & 2.3 & 10.8 & 3.8 & 1.3 & 38 & ``M,'' south in ``W Loop'' \\
12 & 12.2 & 1.8 & 29. & 5.7 & 0.70 & 38 & ``S''+``T,'' in ``W Loop'' \\
13 & 15.0 & 1.4 & 56. & 0.0 & 0.25 & 38 & ``R,'' in ``W Loop'' \\
14 & 4.0 & 0.4 & 32. & 5.3 & 0.19 & 38 & ``J,'' N part of inner bulge \\
15 & 15.0 & 0.8 & 39. & 8.7 & 0.23 & 38 & ``K,'' S part of inner bulge \& nucl.\ of 38 \\
16 & 9.5 & 0.6 & 1.6 & 0.65 & 3.1 & 38 & ``L,'' outer bulge of 38 \\
17 & 5.8 & 0.2 & 0.22 & 3.3 & 4.5 & \nodata & ``Overlap region,'' most dusty part \\
\enddata

\tablenotetext{(a)}{Values for Region 1 based only on eastern part
(western covered by WFPC2 to $<50\%$)}

\end{deluxetable}

\clearpage


\begin{deluxetable}{lcccccccc}
\tabletypesize{\scriptsize}
\rotate
\tablecaption{Number of SN II and Time Required to Produce the Observed Metal 
Masses. 
\label{snmetal}}
\tablewidth{0pt}
\tablehead{
\colhead{\begin{tabular}{c}
Reference\tablenotemark{(*)}\\
(1)
\end{tabular}} &
\colhead{\begin{tabular}{c}
$N_{\rm SNII}$(Ne)\\
$(\times10^5)$\\
(2)
\end{tabular}} &
\colhead{\begin{tabular}{c}
$N_{\rm SNII}$(Mg)\\
$(\times10^5)$\\
(3)
\end{tabular}} &
\colhead{\begin{tabular}{c}
$N_{\rm SNII}$(Si)\\
$(\times10^5)$\\
(4)
\end{tabular}} &
\colhead{\begin{tabular}{c}
$N_{\rm SNII}$(Fe)\\
$(\times10^5)$\\
(5)
\end{tabular}} &
\colhead{\begin{tabular}{c}
$t_{\rm SNII}$(Ne)\tablenotemark{(+)}\\
$(Myr)$\\
(6)
\end{tabular}} &
\colhead{\begin{tabular}{c}
$t_{\rm SNII}$(Mg)\tablenotemark{(+)}\\
$(Myr)$\\
(7)
\end{tabular}} &
\colhead{\begin{tabular}{c}
$t_{\rm SNII}$(Si)\tablenotemark{(+)}\\
$(Myr)$\\
(8)
\end{tabular}} &
\colhead{\begin{tabular}{c}
$t_{\rm SNII}$(Fe)\tablenotemark{(+)}\\
$(Myr)$\\
(9)
\end{tabular}} 
}
\startdata
A96 & $3.4$ & $2.4$ & \nodata & $2.5$ & $11.3$, $1.3$ & $8.0$, $0.9$ & \nodata & $8.4$, $1.0$ \\
T95 & $1.5$ & $1.1$ & $2.0$ & $1.5$ & $4.9$, $0.6$ & $3.7$, $0.4$ & $6.6$, $0.8$ & $4.9$, $0.6$ \\
T95+M92 & \nodata & \nodata & \nodata & $1.5$ & \nodata & \nodata & \nodata & $4.9$, $0.6$ \\
W95;A;$10^{-4}Z_\sun$ & $3.6$ & $3.6$ & $2.5$ & $2.5$ & $12.0$, $1.4$ & $12.0$, $1.4$ & $8.4$, $1.0$ & $8.2$, $0.9$ \\
W95;B;$10^{-4}Z_\sun$ & $1.5$ & $2.0$ & $2.2$ & $2.1$ & $5.1$, $0.6$ & $6.6$, $0.8$ & $7.4$, $0.9$ & $7.0$, $0.8$ \\
W95;A;$Z_\sun$ & $1.9$ & $2.0$ & $2.1$ & $1.6$ & $6.3$, $0.7$ & $6.7$, $0.8$ & $7.0$, $0.8$ & $5.3$, $0.6$ \\
W95;B;$Z_\sun$ & $1.3$ & $1.4$ & $1.8$ & $1.3$ & $4.3$, $0.5$ & $4.6$, $0.5$ & $6.1$, $0.7$ & $4.2$, $0.5$ \\
N98S1 & $1.5$ & $1.1$ & $1.9$ & $2.1$ & $5.0$, $0.6$ & $3.5$, $0.4$ & $6.3$, $0.7$ & $6.9$, $0.8$ \\
N98A1 & $1.5$ & $1.0$ & $2.5$ & $2.1$ & $4.9$, $0.6$ & $3.4$, $0.4$ & $8.2$, $1.0$ & $6.9$, $0.8$ \\
N98A3 & $1.5$ & $1.1$ & $2.7$ & $2.0$ & $4.9$, $0.6$ & $3.5$, $0.4$ & $8.9$, $1.0$ & $6.8$, $0.8$ \\
\enddata

\tablenotetext{(*)}{A96, Arnett 1996; T95, Tsujimoto et al.\ 1995; M92, Maeder 
1992; W95, Woosley \& Weaver 1995; N98, Nagataki et al.\ 1998. N98S1, N98A1, 
and N98A3 are the models of the spherical explosion and the axisymmetric 
explosions, respectively. The degree of deviation from spherical symmetry is 
larger in N98A3 than in N98A1.}
\tablenotetext{(+)}{Computed for a SN rate of 0.03 yr$^{-1}$ and of 0.26 yr$^{-1}$, 
respectively (see text).}

\end{deluxetable}

\clearpage


\begin{deluxetable}{ccccc}
\tablecaption{Thermal energy supplies for the five regions presenting
overabundances of metals. 
\label{eth}}
\tablewidth{0pt}
\tablehead{
\colhead{\begin{tabular}{c}
Reg \#\\
(1)
\end{tabular}} &
\colhead{\begin{tabular}{c}
$E_{\rm th}$(Ne)\\
(erg)\\
(2)
\end{tabular}} &
\colhead{\begin{tabular}{c}
$E_{\rm th}$(Mg)\\
(erg)\\
(3)
\end{tabular}} &
\colhead{\begin{tabular}{c}
$E_{\rm th}$(Si)\\
(erg)\\
(4)
\end{tabular}} &
\colhead{\begin{tabular}{c}
$E_{\rm th}$(Fe)\\
(erg)\\
(5)
\end{tabular}} 
}
\startdata
4b &  $1.02_{-0.74}^{+1.20}\times10^{55}$    &  $9.42_{-6.06}^{+28.44}\times10^{54}$ & $5.41_{-3.69}^{+5.10}\times10^{54}$   & $1.27_{-0.51}^{+2.78}\times10^{55}$  \\
5  &  $1.64_{-0.85}^{+2.51}\times10^{55}$    &  $3.06_{-2.49}^{+51.99}\times10^{55}$ & $1.20_{-1.20}^{+5.62}\times10^{56}$   & $1.02_{-0.35}^{+32.34}\times10^{55}$ \\
7  &  $1.72_{-1.46}^{+3.07}\times10^{55}$    &  $1.23_{-0.84}^{+2.11}\times10^{55}$  & $1.19_{-0.77}^{+1.32}\times10^{55}$   & $6.61_{-2.65}^{+6.32}\times10^{54}$  \\
8a &  $1.30_{-0.82}^{+1.61}\times10^{55}$    &  $6.16_{-4.41}^{+7.33}\times10^{54}$  & $2.50_{-2.28}^{+4.04}\times10^{54}$   & $1.18_{-0.33}^{+0.54}\times10^{55}$  \\
8b &  $5.61_{-5.27}^{+\infty}\times10^{55}$  &  $1.87_{-1.67}^{+46.04}\times10^{55}$ & $2.66_{-2.59}^{+\infty}\times10^{55}$ & $3.61_{-3.15}^{+46.01}\times10^{55}$ \\
\enddata

\end{deluxetable}

\end{document}